\def\cobold{CO$^{\sf 5}$\-BOLD}

\documentclass[
    ,final            
  ]
  {aipproc}

\layoutstyle{6x9}
\pdfoutput=1

\begin{document}

\vspace*{-25mm} \noindent In \emph{Kodai School on Solar Physics}, 
held from Dec.~10--22, 2006 at Kodaikanal Solar Observatory, India, 
S.S.~Hasan and D.~Banerjee (eds.), 2007 American Institute of Physics
Conference Proceedings, Vol.~919
\vspace*{5mm}

\title{Photospheric processes and magnetic flux tubes}


\author{Oskar Steiner}{
  address={Kiepenheuer Institut f\"ur Sonnenphysik, Sch\"oneckstrasse 6,
           D-79104, Freiburg, Germany}
}

\begin{abstract}
New high-resolution observations reveal that small-scale magnetic flux 
concentrations have a delicate substructure on a spatial scale of $0.1''$. Their 
basic structure can be interpreted in terms of a magnetic flux sheet or tube that  
vertically extends through the ambient weak-field or field-free atmosphere
with which it is in mechanical equilibrium. A more refined interpretation 
comes from new three-dimensional magnetohydrodynamic simulations 
that are capable of reproducing the corrugated shape of magnetic flux 
concentrations and their signature in the visible continuum. Faculae are
another manifestation of small-scale magnetic flux concentrations. It is 
shown that the characteristic asymmetric shape of the contrast profile of faculae
is an effect of radiative transfer across the rarefied atmosphere of the 
magnetic flux concentration. Also discussed are three-dimensional radiation 
magnetohydrodynamic simulations of the integral layers from the top of the 
convection zone to the mid-chromosphere. They show a highly dynamic 
chromospheric magnetic field, marked by rapidly moving filaments of stronger 
than average magnetic field  that form in the compression zone downstream 
and along propagating shock fronts. The simulations confirm the picture of 
flux concentrations that strongly expand through the photosphere into a more 
homogeneous, space filling chromospheric field. Future directions in the 
simulation of small-scale magnetic fields are indicated with a few examples 
from recent reports.

The second part of these lecture notes is devoted to a few
basic properties of magnetic flux tubes that can be considered
to be an abstraction of the more complicated flux concentrations known
from observations and numerical simulations. By analytical
means we will find that an electrical current flows in a sheet
at the surface of a flux-tube for which location we also derive the 
mechanical equilibrium condition. The equations for constructing a 
magnetohydrostatic flux tube embedded in a gravitationally stratified 
atmosphere are derived. It is shown that the expansion of a flux tube
with height sensibly depends on the difference in the thermal
structure between the atmosphere of the flux tube and the surrounding
atmosphere. Furthermore, we will find that radiative equilibrium produces a 
smaller temperature gradient within the flux tube compared to that in
the surrounding atmosphere. The condition for interchange stability
is derived and it is shown that small-scale magnetic flux concentrations
are liable to the interchange instability.
\end{abstract}

\maketitle



\section{Part I: The simulation of small-scale magnetic fields}
\label{sect_int_part1}

With ``realistic simulations'' computational physicists aim  at imitating 
real physical processes that occur in nature. In the course of rebuilding 
nature in the computer, they aspire to a deeper understanding of the 
process under investigation. In some sense the opposite approach is taken
by computational physicists that aim at separating the fundamental
physical processes by abstraction from the particulars for obtaining
``ideal simulations'' or an analytical model of the essential physical process. 
Both strategies are needed and are complementary. In the first part
of these lecture notes, we mainly focus on ``realistic simulations'' and 
comparison with observations and provide a more abstract background
in the second part.

The term small-scale flux concentration is used here to designate the 
magnetic field that appears in G-band filtergrams as bright tiny objects 
within and at vortices of intergranular lanes.  They are also visible in 
the continuum, where they are called \emph{facular points} \citep{mehltretter74},
while the structure made up of bright elements is known as the 
\emph{filigree} \citep{dunn+zirker73}. In more recent times, the 
small-scale magnetic field was mostly observed in the G band 
(a technique originally introduced by \citet{muller1985})
because the molecular band-head of CH that constitutes the G band 
acts as a leverage for the intensity contrast 
\citep{rutten99,rutten+al01,sanchez-almeida+al01,shelyag+al04,steiner+al01}.
Being located in the blue part of the visible spectrum, this 
choice also helps improving the diffraction limited spatial resolution 
and the contrast in the continuum. 

Small-scale magnetic flux concentrations are studied for
several reasons:
\begin{list}{-}{\setlength{\topsep}{4pt}\setlength{\parsep}{0.0mm}
                      \setlength{\itemsep}{4pt}\setlength{\leftmargin}{10pt}}
\item[-] Since they make up the small end of a hierarchy of magnetic structures 
         on the solar surface, the question arises whether they
         are ``elemental'' or whether yet smaller flux elements exist.
         How do they form? Are they a surface phenomenon?
         What is their origin?
\item[-] Near the solar limb they can be identified with faculae, known to
         critically contribute to the solar irradiance variation.
\item[-] They probably play a vital role in the transport of mechanical energy 
         to the outer atmosphere, e.g., by guiding and converting magnetoacoustic 
         waves generated by the convective motion and granular buffeting or,
         more directly, by ohmic dissipation.
\end{list}


\section{Structure of small-scale magnetic flux concentrations}
\label{sect_obs}

\begin{figure}
\centering

\begin{minipage}[b]{0.475\textwidth}
\includegraphics[width=1.0\linewidth]{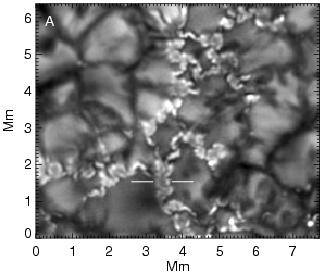}
\end{minipage}\hspace*{0.04\textwidth}
\begin{minipage}[b]{0.475\textwidth}
\includegraphics[width=1.0\textwidth]{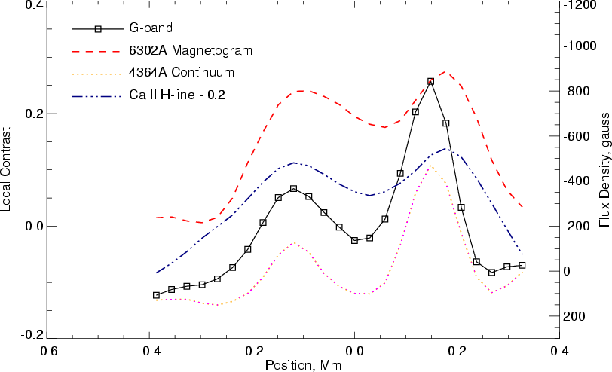}
\vspace*{0.08\linewidth}
\end{minipage}
\caption{\emph{Left:} G-band filtergram showing the ribbon-like shape
of magnetic flux concentrations. \emph{Right:} G-band intensity (solid curve) 
along the section indicated by the horizontal white lines in
the image to the left. Also shown are the magnetogram signal, the
continuum intensity at 436.4~nm, and the Ca II H-line intensity 
down-shifted by 0.2. From \citet{berger+al04}.
\label{steiner_fig01}
}
\end{figure}

Recent observations of unprecedented spatial resolution with the 
1~m Swedish Solar Telescope by \cite{berger+al04} and \cite{vandervoort05}
reveal  G-band brightenings in an active region as delicate, corrugated 
ribbons  that show structure down to the resolution capability of the instrument
of $0.1''$, while isolated point-like brightenings exist as well. The structure made 
up of these objects evolves on a shorter than granular time-scale, giving the 
impression of a separate (magnetic) fluid that resists mixing with the granular 
material. Figure~\ref{steiner_fig01} shows an example G-band filtergram from
the former paper taken in a remnant active region plage near disk center.
In this region, intergranular lanes are often completely filled with 
magnetic field like in the case marked by the white, horizontal lines in 
Fig.~\ref{steiner_fig01}. 
There, and in other similar cases, the magnetic field concentration is framed 
by a striation of bright material, while the central part is dark. 
Besides examples of ribbon bands, Figure~\ref{steiner_fig01} shows also
an isolated bright point in the lower right corner.

The graphic to the right hand side of Fig.~\ref{steiner_fig01} displays the 
emergent G-band intensity (solid curve) from the cross section marked by 
the white horizontal lines in the image to the left. Also shown are the 
corresponding magnetographic signal 
(dashed curve), the blue continuum intensity (dotted), and the Ca H-line
intensity (dash-dotted). We note that the magnetic signal is confined to the
gap between the two horizontal white lines. The intensities show
a two-humped profile.

\begin{figure}
\centering
\includegraphics[width=1.0\linewidth]{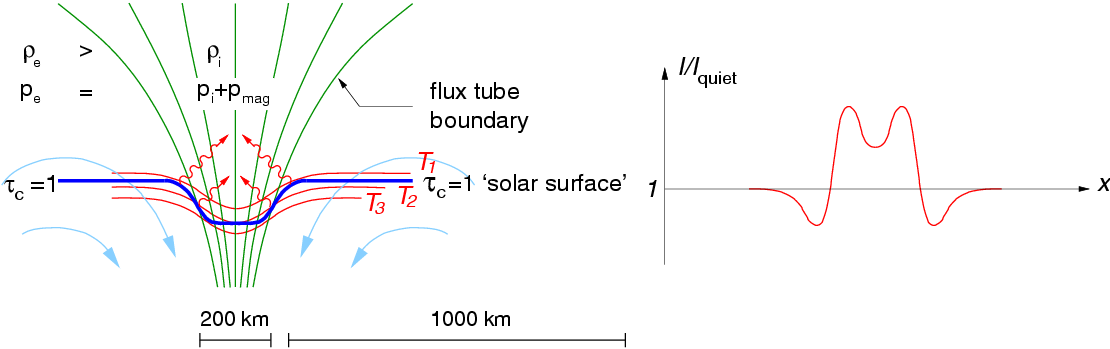}
\caption{Sketch of a magnetic flux sheet (left) with corresponding
intensity contrast (right), distilled from self-consistent numerical
MHD simulations. Note that the isothermal surfaces are not 
exactly parallel to the surface of optical depth unity, which gives rise
to the particular M-shape of the contrast profile. From \citet{steiner07}.
\label{steiner_fig02}
}
\end{figure}

This situation reminds of the flux-sheet model and the ``bright wall effect''
of earlier theoretical work. A first quasi-stationary, self-consistent 
simulation of a small-scale flux sheet was carried out by 
\citet{deinzer+al84a,deinzer+al84b}, popularly known as the ``KGB-models''.
The basic properties of this model are sketched in
Fig.~\ref{steiner_fig02}. (A first kind of this sketch was published by
\citet{zwaan78}.)
Accordingly, a small-scale flux concentration, either 
of tube or sheet-like shape, is in mechanical equilibrium with the ambient
atmosphere, viz., the gas plus magnetic pressure of the atmosphere within the 
tube/sheet balances the gas pressure in the ambient (field-free) medium. 
This calls for a reduced
density in the flux concentration with respect to 
the environment, at least in the photospheric part, where the radiative heat 
exchange quickly drives the configuration towards radiative equilibrium, hence 
to a similar temperature at constant geometrical height. The density reduction 
renders the flux tube/sheet atmosphere  more 
transparent, which causes a depression of the surface of constant  optical depth,
as indicated by the surface of $\tau_c = 1$ in Fig.~\ref{steiner_fig02}. In a plage or
network region, this effect increases the ``roughness'' of the solar surface,
hence the effective surface from which radiation can escape, which increases
the net radiative loss from these areas. 

The graphics to the right hand side of Fig.~\ref{steiner_fig02}  shows a sketch of
the relative intensity emerging from this model, viz., the intensity of light 
propagating in the vertical direction as a function of distance from the  flux sheet's plane of symmetry.
It corresponds to the plot on the right hand side of Fig.~\ref{steiner_fig01}.
The similarity between this model  and  the observation is striking.  
Turning to a narrower flux sheet/tube would result in the merging of the two contrast 
peaks to a single central peak in both, model and observation, i.e., to a ribbon
band or bright point, respectively.
Yet, the striation of the depression wall that can be seen in the observation 
is of course not reproduced by the model, which is strictly two-dimensional with 
translational invariance in lane direction. We will see in the section on the physics
of faculae that three-dimensional magnetoconvection
simulations show a rudimentary striation. The physical origin of the striation is still 
unknown.

Accordingly, the basic properties of ribbon-like magnetic flux concentrations 
can be understood in terms of a magnetic flux sheet embedded in and in force 
balance with a more or less field-free ambient medium. This can also be said
(replacing the word sheet by tube) of the rosette structure visible in other
still images of \citet{berger+al04} who call it ``flower-like''. Flowers can transmute 
to pores and vice versa. The striation of their bright collar is similar
to that seen in ribbon structures. Discarding the striation, the basic 
properties of flowers can well be interpreted in terms of a tube shaped 
flux concentration like the one sketched in Fig.~\ref{steiner_fig02} or
shown in Fig.~\ref{fig_ftradeq}.

A 2~h sequence of images by \citet{vandervoort05} with a quality 
comparable to  Fig.~\ref{steiner_fig01} reveals that the shape of the
ribbon-like flux concentrations and the striation of ribbons and flowers change 
on a very short time-scale, of the order of the Alfv\'en crossing travel time.
This suggests that these morphological changes and the striation itself are
related to the flute instability, which small-scale flux concentrations are liable 
to. For an untwisted axisymmetric flux tube, the radial component of the
magnetic field at the flux-tube surface must decrease with height, 
$\left. {\rm d} B_r /{\rm d} z\right|_S < 0$, 
in order that the flux tube
is stable against the flute (interchange) instability \citep{meyer+al77}.
While sunspots and pores with a magnetic flux in excess of 
$\Phi \approx 10^{11}$\,Wb do meet this condition, small-scale flux concentrations 
do not fulfill it \citep{schuessler84,steiner_phd,buente+al93}. 
\citet{buente93} shows that small-scale flux sheets too, are flute unstable,
and he concludes that filament formation due to the flute instability close
to the surface of optical depth unity would ensue. As the flux sheet is bound 
to fall apart because of the flute instability, its debris are 
again reassembled by the continuous advection of strands of magnetic
field back to the intergranular lane so that a competition between the 
two effects is expected to take place, which might be at the origin of the 
corrugation of the field concentrations and of the striation of the 
tube/sheet interface with the ambient medium. For a derivation of
the flute-instability condition see part II of these lecture notes.

Although the fine structure of small-scale magnetic flux concentrations 
changes on a very short time scale, single flux elements seem to persist
over the full duration of the time sequence of 2 h. They may dissolve
or disappear for a short period of time, but it seems that the same magnetic
flux continually reassembles to make them reappear nearby. Latest G-band 
time sequences obtained with the Solar Optical Telescope (SOT) on board of the
Japanese space satellite HINODE ({\tt http://solarb.msfc.nasa.gov})
confirm these findings even for G-band bright points of low intensity.
This suggest a deep anchoring of at least some of the flux elements
although numerical simulation seem not to confirm this conjecture.

As indicated in the sketch of Fig.~\ref{steiner_fig02}, the magnetic 
flux concentration is framed by a downflow of material, fed by a horizontal 
flow that impinges on the flux concentration. Already the flux-sheet model 
of \citet{deinzer+al84b} showed a persistent flow of this kind. According to
these authors it is due to radiative cooling from the depression 
walls of the magnetic flux concentrations (the ``hot wall effect'') that  causes
a horizontal pressure gradient, which drives the flow. The non-stationary
flux-sheet simulations of \citet{steiner+al1998} and \citet{leka+steiner2001} 
showed a similar persistent downflow, which becomes faster and narrower , 
with increasing depth, turning into  veritable {\it downflow jets} beneath the visible
surface. While downflows in the periphery of pores have been observed
earlier \citep{leka+steiner2001,sanka+rimmele03,tritschler+al03}
and also horizontal motions towards a pore \citep{dorotovic+al02},
only very recently such an accelerating downflow has been observationally 
detected in the immediate vicinity of ribbon bands by \citet{langangen+al07}.


\section{Simulation of small-scale magnetic flux concentrations}
\label{sect_sim}

For a realistic simulation of small-scale magnetic flux concentrations in
the solar photosphere one must solve the system of the 
magnetohydrodynamic equations including continuum, momentum,
induction, and energy equation, preferably in three spatial dimensions. 
Since flux concentrations are
predominantly governed by convective motion, at least the surface
layers of the convection zone must be taken into account in order
to obtain reliable results for the magnetic field in the photosphere.
Hence, the computational domain must span the distinct radial section
of the sun, where energy transport changes from convective to
radiative so that radiative transfer must be taken into account in
the energy equation. An equation of state and opacities appropriate 
for the solar plasma in the region of interest are essential
for a realistic simulation. The spatial extent of the computational domain 
is limited by the computational resources and the minimally required
spatial resolution. Since we would like to resolve at least magnetic flux 
concentrations with a size of 100~km we should have a grid constant not
surpassing about 10~km in the horizontal direction. In the vertical direction 
the resolution can be adjusted according to the variation of the pressure scale 
height but should also be at least $\approx 10$~km in the photosphere. Considering
that a computational grid with $1000^3$ grid cells is definitely the 
uppermost limit for present day simulations, we obtain a size limit for the 
computational domain of about $(10~{\rm Mm})^3$.

\begin{figure}
  \includegraphics[width=0.5\textwidth]{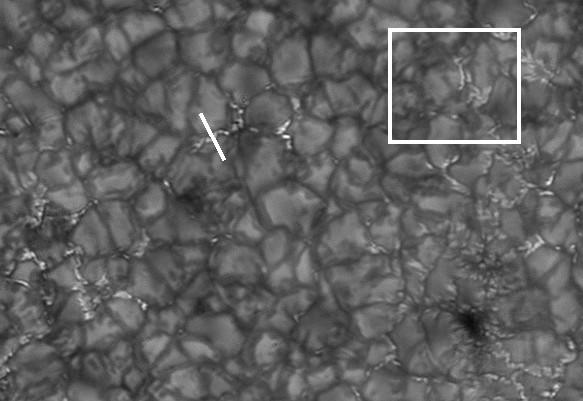}
  \caption{G-band filtergram of an area of $32\times 22$ arcs
  of the solar surface. The white line drawn across an elongated bright 
  ribbon band corresponds to the width of a two-dimensional simulation 
  domain of 2400~km. The white box shows the size of a three-dimensional 
  simulation domain of $4800\times 4800$~km. Image by K.~Mikurda, 
  F.~W\"oger, \& O.~von der L\"uhe with the Vacuum Tower Telescope (VTT) 
  at Tenerife.
  }
  \label{fig_slab_box}
\end{figure}

\begin{figure}
\begin{minipage}[c]{0.45\textwidth}
  \includegraphics[width=1.0\textwidth]{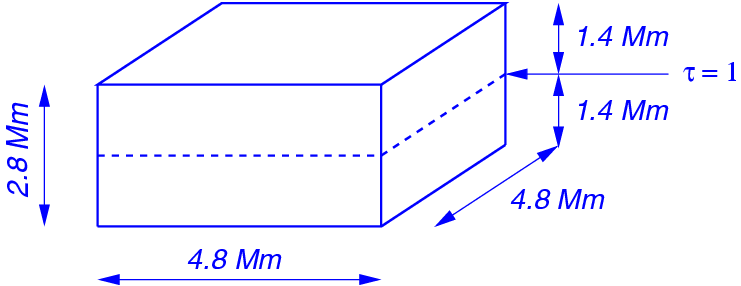}
\end{minipage}\hspace*{-0.10\textwidth}
\begin{minipage}[c]{0.55\textwidth}
  \includegraphics[width=1.0\textwidth]{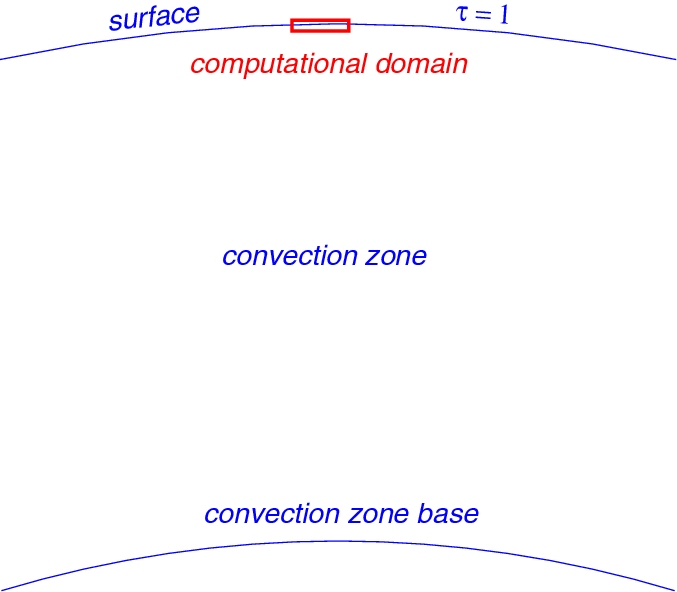}
\end{minipage}\\
\caption{Typical size of a three-dimensional computational box (left) on scale with 
  the convection zone boundaries of the sun (right).
}
\label{fig_boxinthesun}
\end{figure}

Fig.~\ref{fig_slab_box} shows the typical size of the horizontal extent
of a two-dimensional and a three-dimensional computational domain
as used in the past, in relation to granulation and G-band bright points.
We should be aware that such a computational box, despite the large number
of grid cells, encompasses a tiny portion, only, of the solar convection zone. 
This is illustrated in Fig.~\ref{fig_boxinthesun}, where the computational
box of the size given to the left, is shown in relation to the size of
the solar convection zone. Yet, this tiny computational box includes
the full height range of the photosphere and the mass density varies
by about a factor of $10^5$. Even though the box reaches only 1.4~Mm
into the convection zone, it probes a depth range of already about 5 pressure 
scale heights of the convective layers and hydrogen changes from completely 
neutral to a substantial fraction being ionized.

The ideal MHD-equations can be written in conservative form as:
\begin{equation}
  \label{MHD-equations}
  \frac{\partial \mathbf{U}}{\partial t}+\mathbf{\nabla}\cdot
  \mathbf{\mathcal{F}}= \mathbf{S}\,,
\end{equation}
where the vector of conserved variables, 
$\mathbf{U}$, 
the source vector,
$\mathbf{S}$, 
due to gravity and radiation, and the flux tensor, 
$\mathbf{\mathcal{F}}$, are
\begin{equation}
\label{conservative variables}
\mathbf{U}=\left(
\rho, \rho \mathbf{v}, \mathbf{B}, E \right)\,,
\qquad
\mathbf{S}=\left(
0, \rho \mathbf{g}, 0, \rho \mathbf{g}\cdot\mathbf{v}
+ q_{\mathrm{rad}}\right)\,,
\end{equation}
\begin{equation}
\label{eqn_flux_tensor}
\mathbf{\mathcal{F}}=\left(\begin{array}{c}
\rho \mathbf{v} \\[1ex] \rho\mathbf{vv} +
\left(p+\displaystyle\frac{\mathbf{B}\cdot\mathbf{B}}
     {2\mu}\right) \mathbf{I} -
\displaystyle\frac{\mathbf{BB}}{\mu} \\[2ex]
\mathbf{Bv} - \mathbf{vB}\\[1ex]
\left(E+p+\displaystyle \frac{\mathbf{B}\cdot\mathbf{B}}
    {2\mu}\right)\mathbf{v} -
\displaystyle\frac{1}{\mu}\left(\mathbf{v}
    \cdot\mathbf{B}\right)\mathbf{B}
\end{array}\right)\,,
\end{equation}
and where the dyadic tensor product of two vectors 
$\mathbf{a}$ and
$\mathbf{b}$ is the tensor 
$\mathbf{ab} = \mathbf{C}$ with elements
$c_{mn} = a_m b_n$. $\mathbf{v}$ and $\mathbf{B}$
are vectors of the velocity and magnetic fields.
$\mu$ is the magnetic permeability, which can be taken to be
$\mu = \mu_0 = 4\pi\cdot 10^{-7}\,\mbox{[V\,s\,A$^{-1}$\,m$^{-1}$]}$
for the solar plasma. $\mathbf{I}$ is the identity matrix and
$\mathbf{a}\cdot\mathbf{b} = \sum_k a_k b_k$ the scalar product
of the two vectors $\mathbf{a}$ and $\mathbf{b}$.

The total energy $E$ is given by
\begin{equation}
\label{total energy}
E=\rho\epsilon + \rho\frac{\mathbf{v}\cdot\mathbf{v}}{2} + 
\frac{\mathbf{B}\cdot\mathbf{B}}{2\mu},
\end{equation}
where $\epsilon$ is the thermal energy per unit mass.
The additional solenoidality constraint,
\begin{equation}
\label{no monopoles}
\mathbf{\nabla}\cdot \mathbf{B}=0,
\end{equation}
must also be fulfilled.
The MHD equations must be closed by an equation of state which gives the 
gas pressure, $p$, as a function
of the density and the thermal energy per unit mass
\begin{equation}
\label{eos}
p=p(\rho,\epsilon)\,,
\end{equation}
usually available to the program in tabulated form.

The radiative source term is given by
\begin{equation}
q_{\mathrm{rad}} = 4\pi\rho\int \!\!\kappa_{\nu} (J_{\nu} - B_{\nu}){\rm d}\nu\,,
\end{equation}
where $J_{\nu}$ is the mean intensity and $B_{\nu}$ the Planck function. Assuming
strict local thermodynamic equilibrium (LTE), the intensity is given by the
formal solution of the radiative transfer equation:
\begin{equation}
  J_{\nu} (\mathbf{r}) = \frac{1}{4\pi}\oint I_{\nu} (\mathbf{r},
  \mathbf{n}){\rm d}\Omega\,,\qquad
%
  I_{\nu}(\mathbf{r},\mathbf{n}) =  I_{\nu\,0} {\rm e}^{-\tau_{\nu\,0}} 
  + \int_0^{\tau_{\nu\,0}} 
  B_{\nu}(\tau_{\nu}) {\rm e}^{-\tau_{\nu}} {\rm d}\tau_{\nu}\;,
  \label{eqn_formalsolution}
\end{equation}
where $\tau_{\nu\,0}$ is the optical depth at frequency $\nu$ from the boundary 
to the location 
$\mathbf{r}$ along the direction 
$\mathbf{n}$.
${\rm d}\tau = \kappa\, {\rm d}s$, where $\kappa$ is the total 
opacity and $s$ the spatial distance along a line through location 
$\mathbf{r}$ in direction 
$\mathbf{n}$.
Formally, this can be written more compact as
\begin{equation}
  J_{\nu}({\bf r}) = {\Lambda}_{\nu}
  ({\bf r}, {\bf r^{\prime}}) B_{\nu}({\bf r^{\prime}})
  + G_{\nu}({\bf r})\;,
\end{equation}
where ${\Lambda}_{\nu}$ is the integral operator
which adds the intensities at 
$\mathbf{r}$ 
caused by emission
at all the points 
$\mathbf{r^{\prime}}$ 
in the considered
computational domain, and where $G_{\nu}$ is the transmitted
mean intensity due to the incident radiation field into
this domain. 
In practice the frequency
integration is either replaced by using frequency mean quantities 
(Rosseland mean opacity) or it is approximated by a method of
multiple frequency bands \citep{nordlund82}. For a detailed description of the latter
method see \cite{ludwig_thesis} and \cite{ludwig+al94}.

In practice it is not the ideal MHD-equations that are solved but 
rather some kind of a viscous and resistive form of the equations
with flux tensor
\begin{equation}
\label{eqn_flux_tensor_ni}
%
\mathbf{\mathcal{F}}=\left(\begin{array}{c}
\rho \mathbf{v} \\[1ex] 
\rho\mathbf{vv} + \left(p+\displaystyle\frac{\mathbf{B}
     \cdot\mathbf{B}}
     {2\mu}\right) \mathbf{I} -
     \displaystyle \frac{\mathbf{BB}}{\mu} - 
     \mathbf{R} \\[2ex]
 \mathbf{Bv} - \mathbf{vB}+
     \eta[\nabla\mathbf{B} -
     (\nabla\mathbf{B})^T]\\[1ex]
\left(E+p+\displaystyle\frac{\mathbf{B}\cdot\mathbf{B}}
     {2\mu}\right)\mathbf{v} -
\displaystyle \frac{1}{\mu}\left(\mathbf{v}
     \cdot\mathbf{B}\right)\mathbf{B} +
     \eta (\mathbf{j}  \times \mathbf{B}) - 
     \mathbf{R}\cdot \mathbf{v} + 
     \mathbf{q}^{\mathrm{turb}}
\end{array}\right)\,,
\end{equation}
where 
$\mathbf{R} = \nu\rho [(\mathbf{\nabla}
\mathbf{v}) + (\mathbf{\nabla}\mathbf{v})^T
- (2/3)(\mathbf{\nabla}\cdot \mathbf{v})
\mathbf{I}]$ is the viscous stress tensor, 
$\eta = (\nu/{\rm Pr}_{\mathrm{m}}) = 1/(\mu\sigma)$
the magnetic diffusivity with $\sigma$ being the electric conductivity,
and $\eta (\mathbf{j}\times \mathbf{B}) =  
(\eta/\mu)(\nabla\times \mathbf{B})\times
\mathbf{B}$. 
$\mathbf{R}\cdot \mathbf{v}$ is the tensor product $r_{m\,k} v_k$.
${\rm Pr}_{\mathrm{m}}$ is the magnetic Prandtl number.
$\mathbf{q}^{\mathrm{turb}}$ is a turbulent diffusive heat flux, which 
would typically be proportional to the entropy gradient: 
$\mathbf{q}^{\mathrm{turb}} = -(1/{\rm Pr})\nu\rho T \nabla s$, where 
$\rm Pr$ is the Prandtl number.

Typically, $\nu$ is not taken to be the molecular viscosity coefficient 
but rather some \emph{turbulent} value that takes care of the dissipative
processes that cannot be resolved by the computational grid. Such
\emph{subgrid-scale viscosities} should only act where strong velocity gradients
lead to turbulence. Therefore, they typically depend
on velocity gradients like in the Smagorinsky-type of turbulent viscosity
where
\begin{eqnarray}
   \lefteqn{
   \nu^{\mathrm{t}} = c\left\{2\left[
   \left({\partial v_x\over\partial x} \right)^2 +
   \left({\partial v_y\over\partial y} \right)^2 +
   \left({\partial v_z\over\partial z} \right)^2 \right] +\right.}\nonumber \\
&& \left.\left({\partial v_x\over \partial y} + 
         {\partial v_y\over \partial x}\right)^2 + 
   \left({\partial v_x\over \partial z} + 
         {\partial v_z\over \partial x}\right)^2 + 
   \left({\partial v_y\over \partial z} + 
         {\partial v_z\over \partial y}\right)^2
\right\}^{1/2}\,, 
\end{eqnarray}
and where $c$ is a free parameter. This parameter is normally chosen as small
as possible just in order to keep the numerical integration stable and smooth,
but otherwise having no effect on large scales.

Advanced numerical (high resolution) schemes feature an inherent ``dissipation''
(limiters) that acts similar to the explicit dissipative terms shown in the above 
flux tensor, Eq.~(\ref{eqn_flux_tensor_ni}). This \emph{artificial viscosity} is 
made as small as possible but just large enough in order to keep the numerical 
scheme stable and keep gas pressure and total energy positive. One then
only has to program the ideal equations. Of course, in this case it is 
difficult to quote the actual Reynolds and Prandtl number of the simulation 
because these numbers
change then form grid cell to grid cell depending on the flow. Therefore, for 
some applications, it might be preferable to explicitly include the 
dissipative terms in the equations using constant dissipation coefficients, 
which then allows for well defined dimensionless numbers. However one
integrates the ideal equations on a discrete computational grid, one
is locked with a discretization error that normally assumes a form
similar to the dissipative terms in the non-ideal equations.
More details on computational methods for astrophysical fluid flow
can be found in \citet{leveque+al99}. Other very useful text books
on the numerical solution of the gasdynamics equations are
\citep{toro99,leveque02,laney98}. For corresponding solution methods
of the MHD equations see, e.g., 
\citep{balsara04,brio+wu88,shengtai_li04,pen+al03,powell94,wesenberg_thesis} 
and references therein.

\begin{figure}
  \includegraphics[width=0.42\textwidth]{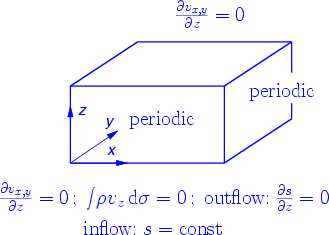}
  \caption{Typical boundary conditions for thermal variables and velocities.
  }
  \label{boundary_cond_therm}
\end{figure}

Fig.~\ref{boundary_cond_therm} displays typical boundary conditions for 
thermal variables and velocity. Noteworthy are the conditions at the bottom 
boundary, where the plasma should freely flow in and out of the computational 
domain (subject to the constraint of mass conservation) since this boundary is 
located within the convection zone. Inflowing material has a given specific
entropy that determines the effective temperature of the radiation leaving
the domain at the top, while the outflowing material carries the entropy it
instantly has. This boundary condition reflects that inflowing material originates 
from the deep convection zone with a close to adiabatic stratification of a fixed 
entropy.

\begin{figure}[h]
\begin{minipage}[c]{0.35\textwidth}
  \includegraphics[width=1.0\textwidth]{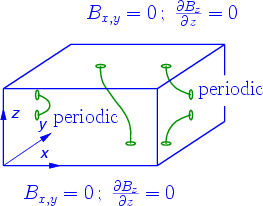}
\end{minipage}
\hspace*{0.05\textwidth}
\begin{minipage}[c]{0.37\textwidth}
  \includegraphics[width=1.0\textwidth]{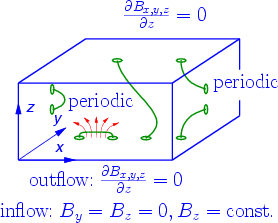}
\end{minipage}
  \caption{Two different realizations for the boundary conditions 
  of the magnetic field. The condition of the right hand panel
  allows for the transport of horizontal magnetic field of a 
  given strength into the box.
  }
\label{fig_boundary_cond_mag}
\end{figure}

Fig.~\ref{fig_boundary_cond_mag} displays two different realizations
(both presently in use) for the boundary conditions of the magnetic 
field. While the magnetic field in the condition of the left
panel is forced to become vertical at the top and bottom boundary,
the condition of the right hand panel allows for the transport
of horizontal magnetic field of a given strength into the box.
More details about this condition are given in the chapter
on future directions. 
%


\subsection{Results from multidimensional simulations}

New results from realistic three-dimensional simulations on the formation, dynamics 
and structure of small-scale magnetic flux concentrations have recently been published 
in a series of papers by Sch\"ussler and collaborators.
\citet{voegler+al05} simulate magnetoconvection in a box encompassing
an area on the solar surface of $6\times 6$~Mm$^2$ with a height extension
of 1400~km, reaching from the temperature minimum to 800~km below the
surface of optical depth unity. Although this is only 
0.4\% of the convection zone depth, the box still includes the 
entire transition from almost completely convective to mainly radiative energy
transfer and the transition from the regime where the flux concentration is dominated 
by the convective plasma flow to layers where the magnetic energy density 
of the flux concentrations by far surpasses the thermal energy density.
The bottom boundary in this and similar simulations is open in the sense
that plasma can freely flow in and out of the computational domain, subject
to the condition of mass conservation. Inflowing material has a given specific
entropy that determines the effective temperature of the radiation leaving
the domain at the top, while the outflowing material carries the entropy it
instantly has.

\begin{figure}
\centering
\includegraphics[width=0.40\linewidth]{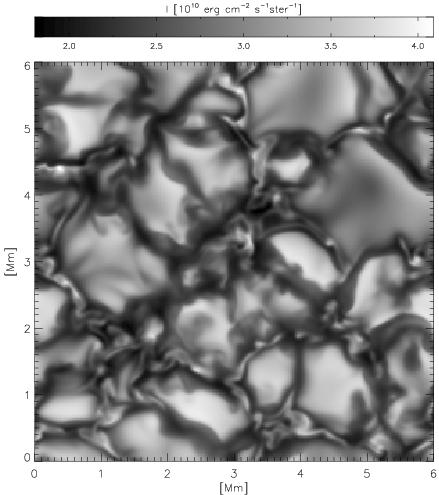}\hspace{0.05\linewidth}
\includegraphics[width=0.40\linewidth]{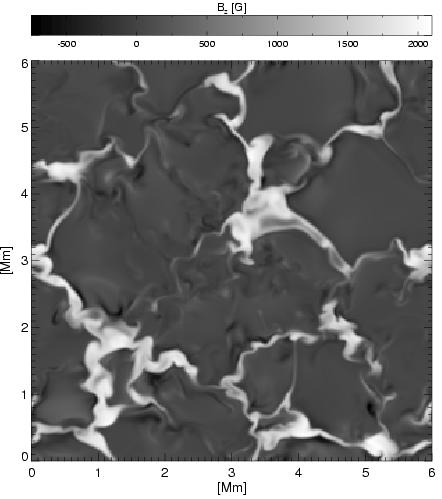}
\caption{Simulation snapshot. \emph{Left:} Frequency integrated intensity. \emph{Right:}
Vertical magnetic flux density at constant average geometrical height of 
optical depth unity. The mean flux density is 0.02~T. From \citet{voegler+al05}.
\label{steiner_fig03}
}
\end{figure}

Figure~\ref{steiner_fig03} shows a snapshot from this simulation: To the left the
emergent mean intensity, to the right the vertical magnetic field strength
at a constant height, viz., at the horizontally averaged geometrical height of 
optical depth unity. (I would like to caution that this magnetic map is not
what would be seen with a magnetograph, irrespective of its spatial resolution,
because it refers to a plane parallel section, which is not what is sampled with
a magnetographic or polarimetric mapping.)
The strong magnetic field in intergranular lanes is manifest
in a corresponding signal in the emergent intensity very much like
the observations  shown in the previous sections (Figs.~\ref{steiner_fig01} and
\ref{fig_slab_box}). Also, the intensity signal
shows the same corrugated and knotted ribbon structure that is observed, and 
sometimes there appear also broader ribbon structures with a dark central core,
like the one marked in Fig.~\ref{steiner_fig01}.
The characteristic striation of the latter case however, is absent in the simulation, 
possibly because the flute instability is suppressed on very small spatial scales 
due to lack of sufficient resolution of the simulation. In the central part of the
snapshot, a micro pore or magnetic knot has formed.

A comparison of the average gas plus magnetic pressure as a function of height
at locations of magnetic flux concentrations with the run of the average gas 
pressure in weak-field regions reveals that the two are almost identical, proving
that even in this dynamic regime, the thin flux tube approximation is very well 
satisfied \citep{voegler+al05}. This result  confirms that the model discussed in 
the previous section and sketched in Fig.~\ref{steiner_fig02} 
is indeed an acceptable first approximation to the real situation.

\begin{figure}[t!]
   \begin{minipage}{1.0\textwidth}
   \centering
   \rotatebox{90}{\includegraphics[height=0.75\textwidth]{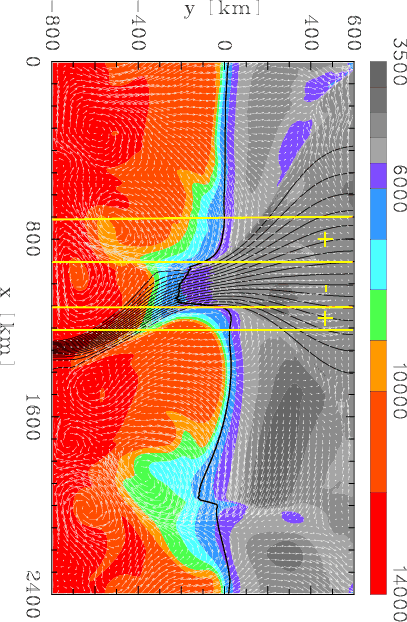}}\\
   \hspace{4mm}\includegraphics[width=0.7\textwidth]{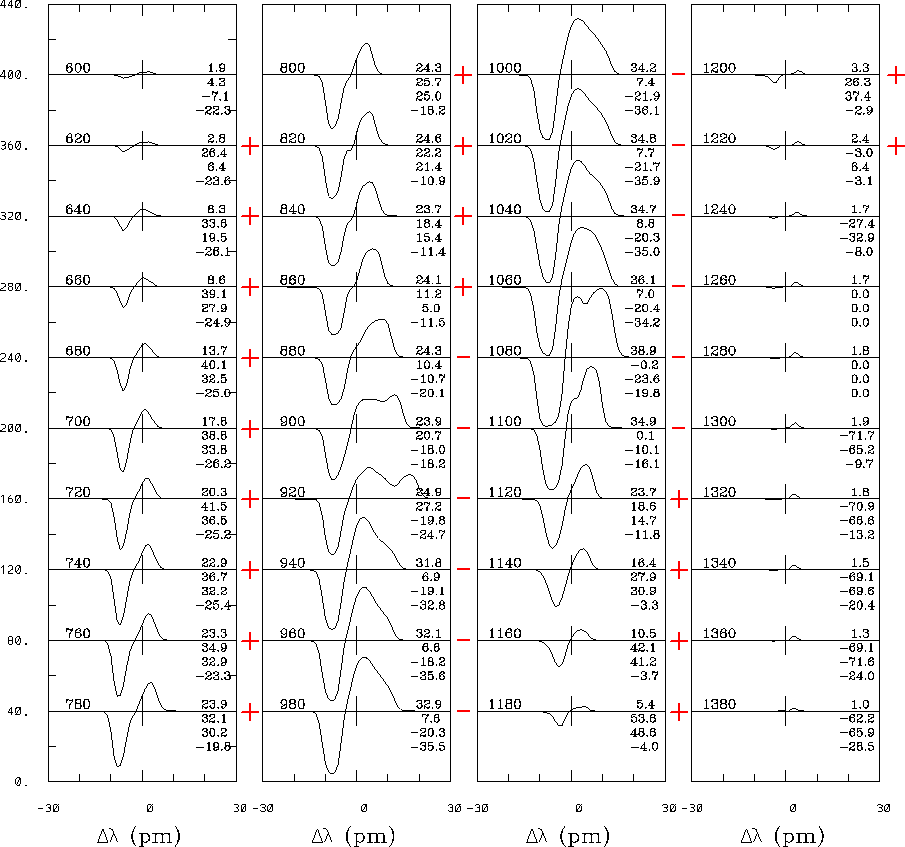}
   \end{minipage}
\caption{\emph{Top:} Snapshot of a two-dimensional simulation of a
magnetic flux sheet in interaction with convective motion. Color coding
indicates temperature in K, white arrows the velocity.  \emph{Bottom:}
Stokes $V$ profiles that emanate
from vertical lines of sight distributed over the horizontal 
interval between 600 and 1380~km of the simulation snapshot. 
Positive area asymmetries result in the two
regions labeled with ``$+$'', while the lines of sight in
the middle region labeled with ``$-$'' contribute negative area 
asymmetries. The actual profiles emanating from
each line of sight is also shown, where the mean amplitude,
the amplitude asymmetry, the area asymmetry (all in \%),
and the zero-crossing shift in m{\AA} are printed (from
top to bottom) on the right hand side of each panel. In addition,
positive and negative area asymmetries are annotated in the right
margin with a plus and a minus sign, respectively.
From \citet{steiner99}.
}
\label{fig_2D_snapshot}
\end{figure}

Simulations are not just carried out for the sake of reproducing observed
quantities. Once good agreement with all kind of observations exists, simulations
allow with some confidence to inform about regions not directly accessible
to observations, for example about the magnetic structure in subsurface
layers. In this respect the simulations of \citet{voegler+al05} show that often 
flux concentrations that have formed at the surface disperse again in shallow depths.
This behaviour was also found by \citet{schaffenberger+al05} in
their simulation with an entirely different code and further by
\citet{stein+nordlund06}.   A vertical section through a 
three-dimensional simulation domain of \citet{schaffenberger+al05}, where two such
shallow flux concentrations have formed, is shown in Fig.~\ref{steiner_fig09}.
The superficial nature of magnetic flux concentrations in the simulations,
however, is difficult to reconcile with the observation that
many flux elements seem to persist over a long time period.

Although the turbulent nature of convection requires simulation in three-dimensional
space, it can be instructive to carry out less costly two-dimensional calculations
that allow for high spatial resolution. For example, a two-dimensional approach 
can be justified for the simulation of elongated sheets of magnetic flux concentrations
as they can be found in intergranular lanes of active regions, e.g., the G-band bright
string that extends perpendicular to the 2-D domain indicated in Fig.~\ref{fig_slab_box}.

Fig.~\ref{fig_2D_snapshot} (top) shows a time instant of a two-dimensional 
simulation of a magnetic flux sheet in interaction with convective motion
\citep{steiner+al1998}. A granule has grown to the right of the flux sheet 
and pushes it to the left. Such ``granular buffeting'' can produce waves 
propagating along, within, and in the environment of a magnetic flux sheet.
Besides a swaying motion of the flux sheet that excites transverse tube
waves, the simulation also revealed that often shock waves propagate
in the vertical direction in the tenuous  upper photospheric part within the 
flux sheet . Such a shock wave is visible in Fig.~\ref{fig_2D_snapshot} (top) in a
height range between 300 to 400~km slightly left of the central part
of the flux sheet. 

From simulations as the ones shown in Figs.~\ref{steiner_fig03} or 
\ref{fig_2D_snapshot} (top),
synthetic observable quantities like intensity contrasts, or spectral lines, 
or polarimetric quantities can be computed and directly compared with real 
observations. For example, Fig.~\ref{fig_2D_snapshot} (bottom)
displays Stokes $V$ profiles computed
for vertical lines of sight distributed over the horizontal 
interval between 600 and 1380~km of the simulation snapshot. 
The shock wave produces the
pathological Stokes $V$ profiles belonging to $x=900$ and $920$~km.
There is a systematic change in the area asymmetry of the Stokes $V$ profiles
across the flux sheet. While the central part of the flux sheet produces
profiles with a negative area asymmetry, the peripheral profiles are positive.
This behaviour can be understood, noting that the flux sheet is
most of the time framed by a strong downflow of material 
\citep{steiner99,leka+steiner2001}.

As mentioned in the first chapter, such a downflow is indeed visible in
high resolution Dopplergrams of small-scale magnetic flux
concentrations \citep{langangen+al07}. Since polarimetric measurements
have not yet reached a similarly good quality in spatial resolution, the corresponding
polarimetric signal of positive Stokes $V$ area asymmetry in the
periphery of flux concentrations has not yet been observationally confirmed, 
but I am confident that it soon will be. 

Other results from the computation of synthetic Stokes profiles from 
three-dimensional magnetohydrodynamic simulations have recently been 
published by \cite{khomenko+al05a} and \cite{shelyag07}. Further recent
three-dimensional magnetohydrodynamic simulations of small-scale
solar magnetic fields include the works \cite{stein+nordlund06} and 
\cite{ustyugov06}.


\section{The physics of faculae}
\label{sect_fac}

With growing distance from disk center, small-scale 
magnetic flux concentrations grow in contrast against the quiet 
Sun background and become apparent as solar faculae close to the limb. 
Ensembles of faculae form plage and network faculae that 
are as conspicuous features of the white light solar disk, 
as are sunspots. It is therefore not surprising that they play a key 
role in the solar irradiance variation over a solar cycle and on 
shorter time scales \citep{fligge+al00,wenzler+al05,foukal+al06}.
Measurements of the center to limb variation of the continuum contrast 
of faculae are diverse,
however, as the contrast is not only a function of  the heliocentric
angle, $\mu = \cos\theta$, but also of facular size, magnetic field
strength, spatial resolution, etc., and as measurements are prone to selection 
effects. While many earlier measurements report a contrast maximum around
$\mu \approx 0.2\ldots 0.4$ with a decline towards the limb, latest measurements
\citep{sutterlin+al99,ahern+chapman00,adjabshirizadeh+koutchmy02,
ortiz+al02,centrone+ermolli03,vogler+al05} point rather to a monotonically 
increasing or at most mildly decreasing contrast out to the limb.

The standard facula model \citep{spruit76}, again consists of a magnetic flux 
concentration embedded in and in mechanical equilibrium with a weak-field 
or field-free environment as is sketched in Fig,~\ref{steiner_fig02}. When
approaching the limb, the limb side of the bright depression wall becomes ever 
more visible and ever more perpendicular to the line of sight, which increases its
brightness compared to the limb darkened environment. At the extreme limb, 
obscuration by the centerward rim of the depression starts to take place, 
which decreases the size and possibly the contrast of the visible limb-side wall.

\begin{figure}
\centering
\begin{minipage}[b]{0.40\textwidth}
\includegraphics[width=1.0\linewidth]{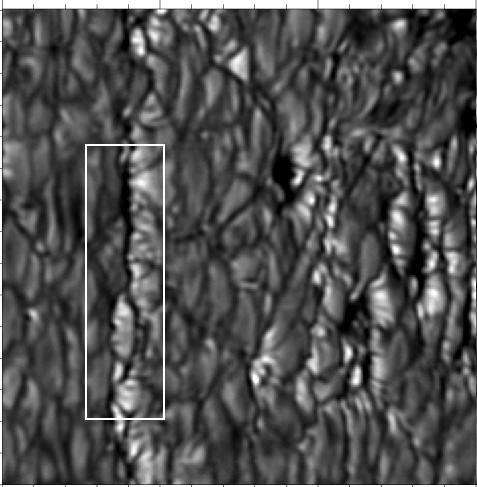}
\end{minipage}\hspace{0.05\linewidth}
\begin{minipage}[b]{0.45\textwidth}
\includegraphics[width=1.0\textwidth]{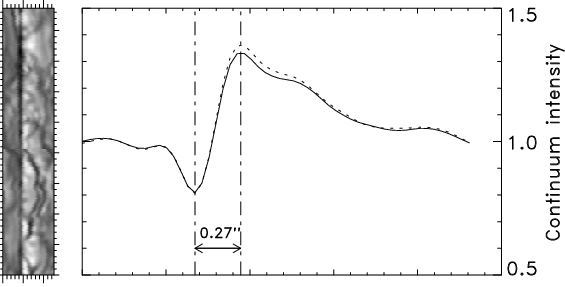}
\vspace*{0.05\linewidth}
\end{minipage}
\caption{\emph{Left:} Network faculae at a heliocentric angle of $\mu = 0.48$ in the 
continuum at 587.5~nm. The solar limb is to the right.
\emph{Right:} Faculae within the white box of the image to the left aligned according 
to the position of the dark lane, together with the mean spatial scan through the 
aligned faculae. The dotted curve refers to profiles for which the maximum intensity
exceeds 1.3.
From \citet{hirzberger+wiehr05}.
\label{steiner_fig04}
}
\end{figure}

Recently, \citet{lites+al04} and \citet{hirzberger+wiehr05} have obtained excellent
images of faculae with the 1~m Swedish Solar Telescope. Figure~\ref{steiner_fig04}
(taken from  \citep{hirzberger+wiehr05}) shows on the left hand side network faculae 
at a heliocentric angle of $\mu = 0.48$ in the continuum at 587.5~nm. 
The solar limb is located 
towards the right hand side. It is clearly visible from this image that faculae are 
in reality partially brightened granules with an exceptionally dark and wide
intergranular lane (``dark facular lane'') on the disk-center 
side of the contrast enhancement, which is also the location of the magnetic flux
concentration. The right half of Fig.~\ref{steiner_fig04} shows the string of faculae
within the white box of the image to the left, aligned according to the position
of the dark lane. Also shown is the mean contrast profile, averaged over the alignment. 
Similar contrast profiles of single faculae are shown by \citet{lites+al04}. Such 
contrast profiles pose now a new constraint that any model of faculae must satisfy.

Magnetoconvective simulations as the one discussed in the previous section
indeed show facular-like contrast enhancements when computing the
emergent intensity along lines of sight that are inclined to the vertical direction
for mimicking limb observation. Such tilted three-dimensional simulation 
boxes are shown in the
papers by \citet{keller+al04}, \citet{carlsson+al04}, and \citet{depontieu+al06}. 
\citet{keller+al04} also show the contrast profile of two isolated ``faculae'', 
which however have a more symmetric shape, rather than the observed characteristic 
steep increase on the disk-center side with the gentle decrease  towards the limb. 
Also they obtain a maximum contrast of 2, far exceeding the observed value of 
about 1.3. It is not clear what the reason for this discrepancy might be. 
Interestingly, the old ``KGB-model'' \citep{deinzer+al84b,knoe+msch88} as well as 
the two-dimensional, non-stationary simulation of \citet{steiner05}
do nicely reproduce the asymmetric shape and the dark lane.

Another conspicuous property of faculae that high-resolution images reveal is that
they are not uniformly bright but show a striation not unlike to and possibly
in connection with the one seen in G-band ribbons at disk center. While this 
feature cannot be reproduced in a two-dimensional model, it must be part of a
satisfactory three-dimensional model. But so far 3-D simulations show only
a rudimentary striation. This finding, rather surprisingly, indicates that the 
effective spatial resolution of present-day three-dimensional simulations is 
inferior to the spatial resolution of best current observations.

\begin{figure}
\centering
\includegraphics[width=1.0\linewidth]{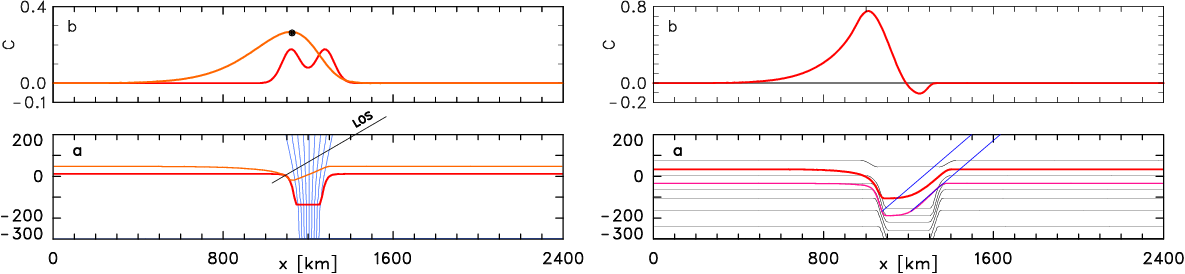}
\caption{
\emph{Left:} a) Magnetic flux concentration (blue, vertically oriented lines of force, 
see internet version for colours) 
with surfaces of optical depth $\tau = 1$ for vertical lines of sight 
(thick/red curve) and lines of sight inclined by $\theta = 60^{\circ}$ to the
vertical direction (thin/red curve). b) Corresponding contrast curves. 
All values of the light red curve left of the black dot originate from lines 
of sight left of the one indicated in panel a).
\emph{Right:}  a) Surfaces of optical depth $\tau = 1$ and $5$ (thick/red) for 
lines of sight inclined by $50^{\circ}$ to the vertical, together with isotherms. 
b) Contrast profile. The region of negative contrast is bounded by the 
two oblique lines of sight indicated in panel a). Adapted from \citet{steiner05}.
\label{steiner_fig05}
}
\end{figure}

\begin{figure}
\centering
\includegraphics[width=0.45\linewidth]{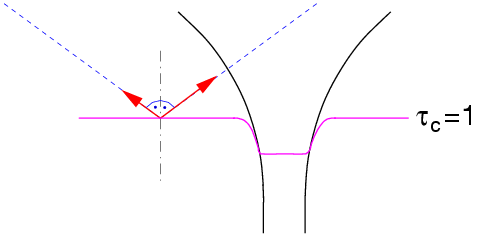}
\caption{Photons preferentially escape along the line of sight to the right that
traverses the magnetic flux sheet/tube in comparison to the line of sight to the
left under equal zenith angle, because of the rarified (less opaque) atmosphere
in the flux sheet/tube. From \citet{steiner07}.
\label{steiner_fig06}
}
\end{figure}

In an attempt to better understand the basic properties of faculae,
\citet{steiner05} considers the ideal model of a magnetohydrostatic flux sheet 
embedded in a plane parallel standard solar atmosphere. For the construction
of this model the flux-sheet atmosphere is first taken to be identical to the 
atmosphere of the ambient medium but shifted in the 
downward direction by the amount of the ``Wilson depression'' (the depression of the 
surface of continuum optical depth unity at the location of the flux concentration).
The shifting results in a flux-tube atmosphere that is less dense and cooler than
the ambient medium at a fixed geometrical height.
In the photospheric part of the flux concentration, thermal equilibrium with the 
ambient medium is then enforced. Denoting with index $i$ the flux-sheet
atmosphere and with $e$ the ambient atmosphere and with $W$ the depth of the
``Wilson depression'', we therefore have
\begin {equation}
T_i(z) = \!\!\left\{ \!\!\! \begin{array}{l}
T_e(z+W)\quad \mbox{for}\quad \tau_c \gg 1\\
T_e(z)\quad \mbox{for}\quad \tau_c \ll 1\;,
\end{array}\right.
\end {equation}
where $\tau_c$ is the optical depth in the visible continuum and
$\rho_i(z) < \rho_e(z)\; \forall z$. The lower left panel
of Fig.~\ref{steiner_fig05} shows this configuration together with two surfaces of
optical depth unity, one for vertical lines of sight (disk center), the other for 
lines of sight running from the top right to the bottom left under an angle of 
$\theta = 60^{\circ}$ to the vertical, like the one indicated in the figure. 
The upper left panel shows the corresponding continuum enhancement for 
disk center (double humped profile) and 
$\theta = 60^{\circ}$. Of the curve belonging to $\theta = 60^{\circ}$, all values
left of the black dot belong to lines of sight left of the one indicated in the lower 
panel. This means that the contrast enhancement extends far beyond the
depression proper in the limbward direction, exactly as is observed. The
reason for this behaviour is explained with the help of Fig.~\ref{steiner_fig06} 
as follows.

A material parcel located in the solar atmosphere and lateral to the flux sheet ``sees'' 
a more  transparent atmosphere in the direction toward the flux sheet compared 
to a direction under equal zenith angle but pointing away from it because of
the rarefied flux-sheet atmosphere. Correspondingly, from a wide area 
surrounding the magnetic flux sheet or flux tube, radiation escapes more 
easily in the direction  towards the flux sheet so that a single flux sheet/tube impacts 
the radiative escape in a cross-sectional area that is much wider than the 
magnetic field  concentration proper.

The right hand side of Fig.~\ref{steiner_fig05} shows a similar situation as to
the left but for a flux sheet that is twice as wide. The continuum contrast
for lines of sight inclined by $\theta = 50^{\circ}$ to the vertical is shown in the
top panel. It can be seen that a dark lane of negative contrast 
 occurs on the disk-center side of the facula.
It arises from the low temperature gradient of the flux-sheet atmosphere in the 
height range of $\tau_c = 1$ and its downshift relative to the external atmosphere in 
combination with the inclined lines of sight. One could say that the dark lane 
in this case is an expression of the cool ``bottom'' of the magnetic flux sheet. 

It is remarkable that this basic, energetically not self-consistent
model is capable of producing both, the facular dark lane and the asymmetrically
shaped contrast curve of the facula, with realistic contrast values. The
results of this basic hydrostatic model carry over to a fully self-consistent
model of a magnetic flux sheet in dynamic interaction with non-stationary 
convective motion \citep{steiner05}. In this case the facular lane becomes 
broader and darker.

It follows from these insights that a facula is not to be identified with 
bright plasma that sticks,  as the name may imply, like a torch out of 
the solar surface and as the ``hillock model'' of \citet{schatten+al86} 
suggests. Rather is it the manifestation of photospheric granulation, seen 
across a magnetic flux concentration --- granulation that appears brighter 
than normal in the form of so called ``facular granules''.
Interestingly, already \citet{chevalier1912} wondered: ``La granulation que l'on 
voit autour des taches plus \'eclatante que sur les autres parties est-elle la 
granulation des facules ou celle de la photosph\`ere vue \`a travers les facules ?''
and \citet{bruggencate40} noted that ``Sie [Photosph\"arengranulen und Fackelgranulen]
unterscheiden sich nicht durch ihre mittlere Gr\"osse, wohl aber durch den Kontrast 
gegen\"uber der Umgebung.''

\begin{figure}
\includegraphics[width=0.48\linewidth]{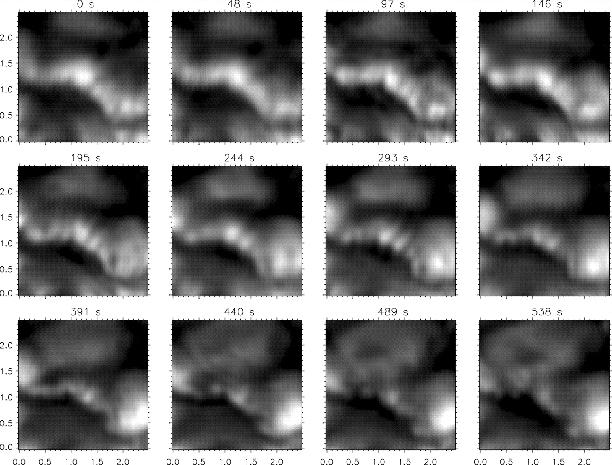}\hspace{0.015\textwidth}
\includegraphics[width=0.48\linewidth]{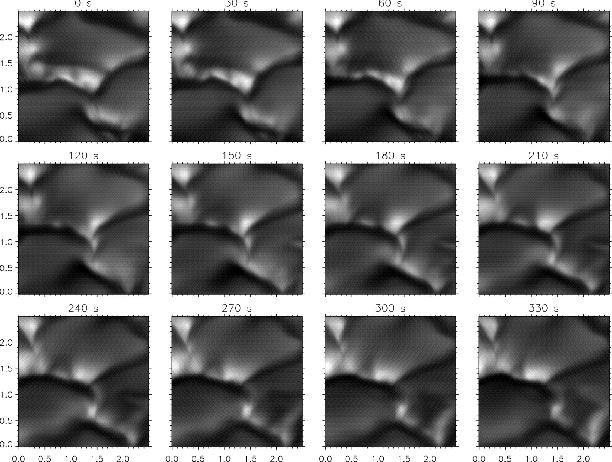}
\caption{\emph{Left,} observation: A dark band gradually moves toward the 
disk center (down) and seemingly sweeps over and "erases" the facula
temporarily. \emph{Right,} simulation: At t = 0~s a dark band is just above 
the center of the figure. It moves toward the disk center (downward), 
and the facula is reduced to a weak band at t = 90~s. 
From \citet{depontieu+al06}.
\label{steiner_fig07}
}
\end{figure}

If this is true, one expects facular granules to show the same dynamic phenomena like
regular granulation. Indeed, this is confirmed in a comparison of observations with
three-dimensional simulations by \citet{depontieu+al06}. 
Fig.~\ref{steiner_fig07} shows on the
left hand side an observed time sequence of a facula in which a dark band gradually moves
from the limb side toward the disk center (down) and seemingly sweeps over and "erases" the 
facula temporarily. The panel on the right hand side shows the same phenomenon in
a time sequence from a three-dimensional simulation. At t = 0~s a dark band is just 
above the center of the figure. It moves toward the disk center (downward in the figure), 
and the faculae are reduced to a weak band at t = 90~s. 
%

Examination of the simulation sequence reveals that dark bands are a consequence of
the evolution of granules. Often granules show a dark lane that usually introduces
fragmentation of the granule. The smaller fragment often dissolves (collapses) in 
which case the dark lane disappears with the collapsing small fragment in the 
intergranular space. Exactly this process can lead to the dark band phenomenon,
when a granular dark lane is swept towards the facular magnetic flux concentration.
Since the facular brightening is seen in the disk-center
facing side of granules, only granular lanes that are advected in the direction of
disk center lead to facular dimming. This example nicely demonstrates how regular 
granular dynamics when seen across the facular magnetic field can lead to 
genuine facular phenomena.
 
Despite the major progress that we have achieved in understanding the physics of 
faculae over the past few years, open questions remain. These concern
\begin{list}{-}{\setlength{\topsep}{3pt}\setlength{\parsep}{0.0mm}
                      \setlength{\itemsep}{3pt}\setlength{\leftmargin}{10pt}}
\item[-] a comprehensive model of the center to limb 
         variation of the brightness of faculae including dependence on size, magnetic flux,
         flux density, color, etc.,
\item[-] a quantitative agreement between simulation and observation with respect to
         measurements in the infrared and with respect to the observed
         geometrical displacement between line core and continuum filtergrams of 
         faculae,
\item[-] the physical origin of the striation, 
\item[-] a quantitative evaluation of the heat leakage caused by faculae, or
\item[-] the role of faculae in guiding magnetoacoustic waves into the chromosphere.
\end{list}


\section{3-D MHD simulation from the convection zone to the chromosphere}
\label{sect_3dmhd}

For investigating the connection between photospheric small scale magnetic fields
and the chromosphere, \citet{schaffenberger+al05} have extended the three-dimensional
radiation hydrodynamics code
\cobold\footnote{www.astro.uu.se/\~{}bf/co5bold\_main.html}
to magnetohydrodynamics for studying
magnetoconvective processes in a three-dimensional environment
that encompasses the integral layers from the top of the convection
zone to the mid chromosphere. The code is based on a finite volume scheme,
where fluxes are computed with an approximate Riemann-solver 
\citep{leveque+al99,toro99} for automatic shock capturing. For the advection 
of the magnetic field components, a constrained transport scheme is used.

\begin{figure}[t!]
\centering
\begin{minipage}[b]{0.70\textwidth}
  \includegraphics[width=1.0\textwidth]{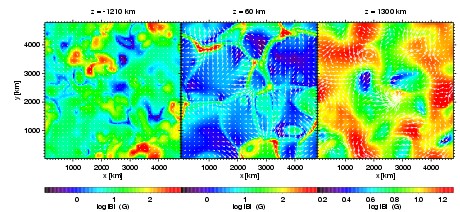}
\end{minipage}\hspace*{0.01\textwidth}
\begin{minipage}[b]{0.275\textwidth}
  \includegraphics[width=1.0\linewidth]{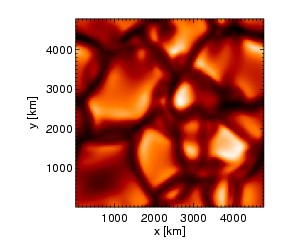}
  \vspace*{0.07\textwidth}
\end{minipage}
\caption{Horizontal sections through the three-dimensional computational domain. 
Color coding displays $\log |B|$ with individual scaling for each panel
(see internet version for colours).  \emph{Left:}
Bottom layer at a depth of 1210~km. \emph{Middle:} Layer 60~km above optical depth 
$\tau_c = 1$. \emph{Right:} Top, chromospheric layer in a height of 1300~km. White 
arrows indicate the horizontal velocity on a common scaling. 
Longest arrows in the panels from left to right correspond to
4.5, 8.8, and 25.2~km/s, respectively. 
\emph{Rightmost:} Emergent visible continuum intensity.
From \citet{schaffenberger+al05}.
\label{steiner_fig08}
}
\end{figure}

\begin{figure}[b!]
\centering
 \includegraphics[width=0.9\linewidth]{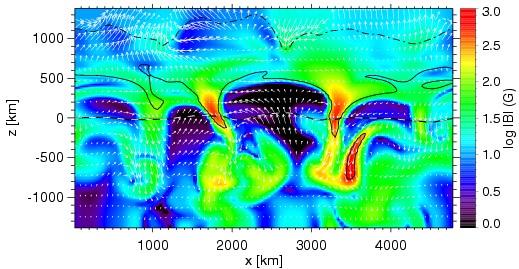}
\caption{Snapshot of a vertical section through the three-dimensional 
computational domain, showing $\log |B|$ (color coded) and velocity 
vectors projected on the vertical plane (white arrows). The b/w dashed 
curve shows optical depth unity and the dot-dashed and 
solid black contours $\beta = 1$ and 100, respectively. 
See internet version for colours. From \citet{schaffenberger+al05}.
\label{steiner_fig09}
}
\end{figure}

The three-dimensional computational domain extends from 1400~km below
the surface of optical depth unity to 1400~km above it and it has a horizontal
dimension of $4800\times 4800$~km. The simulation starts with a homogeneous,
vertical, unipolar magnetic field of a flux density of 1~mT superposed on a 
previously computed, relaxed model of thermal convection. This low flux density 
is representative for magnetoconvection in a very quiet network-cell interior.
The magnetic field is constrained to have vanishing
horizontal components at the top and bottom boundary, but lines of force
can freely move in the horizontal direction, allowing for flux concentrations
to extend right to the boundaries. Because of the top boundary being
located at mid-chromospheric heights, the magnetic field is allowed to freely 
expand with height through the photospheric layers into the more or less
homogeneous chromospheric field.

Figure~\ref{steiner_fig08} shows the logarithmic
absolute magnetic flux density in three horizontal sections through the 
computational domain at a given time instant, together with the emergent
Rosseland mean intensity.
The magnetic field in the chromospheric part is marked by strong dynamics 
with a continuous rearrangement of magnetic flux on a time scale of
less than 1~min, much shorter  than in the photosphere or in the
convection-zone layers. There, the field has a strength between 0.2 and 4~mT
in the snapshot of Fig.~\ref{steiner_fig08}, which is typical for the whole 
time series. Different from the surface magnetic field, it is more 
homogeneous and practically fills the entire space so that the magnetic 
filling factor in the top layer is close to unity. There seems to be no 
spatial correlation between chromospheric flux concentrations and 
the small-scale field concentrations in the photosphere. 

Comparing the flux density of the panel corresponding to $z = 60$~km with 
the emergent intensity, one readily sees that the magnetic
field is concentrated in intergranular lanes and at lane vertices. However,
the field concentrations do not manifest a corresponding intensity signal
like in Fig.~\ref{steiner_fig03} (left). This is because the magnetic flux is too weak 
to form a significant ``Wilson depression'' (as can be seen from 
Fig.~\ref{steiner_fig09}) so that no radiative channeling effect 
(see radiative equilibrium solution in part II) takes place.

Figure~\ref{steiner_fig09} shows the logarithm of the absolute field strength 
through a vertical section of the computational domain. 
Overplotted are white arrows indicating the velocity field. The b/w dashed
curve corresponds to the optical depth unity for vertical lines of
sight. Contours of the ratio of thermal to magnetic pressure, 
$\beta$, for $\beta = 1$ (dot-dashed) and $\beta = 100$ (solid) are also 
shown. Magnetoacoustic waves that form transient filaments of stronger 
than average magnetic field are a ubiquitous phenomenon in the 
chromosphere and are also present in the snapshot of 
Fig.~\ref{steiner_fig09}, e.g., 
along the contour of $\beta = 1$ near $x = 1200$~km and $x = 2500$~km.
They form in the compression zone downstream and along propagating 
shock fronts. These magnetic filaments that have a field strength rarely 
exceeding 4~mT, rapidly move with the shock fronts and quickly form 
and dissolve with them.  
The surface of $\beta = 1$ separates the region of highly dynamic magnetic fields 
around and above it from the more slowly evolving field of high beta plasma below
it. This surface is located at approximately 1000~km but it is corrugated 
and its local height strongly varies in time.

\begin{figure}
\centering
 \includegraphics[width=0.95\linewidth]{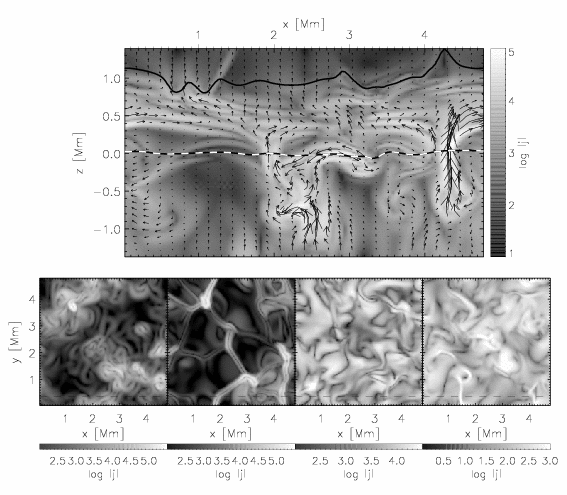}
\caption{Logarithmic current density, $\log |j|$, in a vertical cross 
   section (top panel) and in four horizontal cross sections (bottom panels) in a depth of 
   1180\,km below, and at heights of 90\,km, 610\,km, and 1310\,km above the 
   mean surface of optical depth unity from left to right, respectively.
   The arrows in the top panel indicate the magnetic field strength and
   direction. The dashed line indicates the position of the vertical section.
   [$j$] = $3\,\times\,10^5$ A/m$^{2}$. From \citet{wedemeyer+al07a}.
   \label{steiner_fig10}
}
\end{figure}

A very common phenomenon in this simulation is the formation of a
`magnetic canopy field' that extends in a more or less horizontal
direction over expanding granules and between photospheric flux
concentrations. The formation of such canopy fields proceeds by the
action of the expanding flow above granule centres. This flow
transports `shells' of horizontal magnetic field to the upper
photosphere and lower chromosphere, where shells of different field
directions may be pushed close together, leading to a complicated
network of current sheets in a height range from approximately 400 to
900\,km.

This network can be seen in Fig.~\ref{steiner_fig10} (top), which shows, for
a typical snapshot of the simulation, the logarithmic current density,
$\log |j|$, 
together with arrows indicating the magnetic field strength and direction. 
Figure~\ref{steiner_fig10} (bottom) shows from left to right 
$\log |j|$ 
in four horizontal cross sections in a depth of
1180\,km below, and at heights of 90\,km, 610\,km, and 1310\,km above
the mean surface of optical depth unity. Higher up in the chromosphere
(rightmost panel), thin current sheets form along shock fronts, e.g.,
in the lower left corner near \mbox{$x = 1.4$\,Mm}.

Using molecular values for the electrical conductivity, 
\citet{wedemeyer+al07a} derive an energy flux of 5 to 50\,W\,m$^{-2}$
into the chromosphere caused by ohmic dissipation of these 
current sheets. This value is about two orders of magnitude short 
of being relevant for chromospheric heating. On the other hand,
the employed molecular values for the conductivity might be orders
of magnitude too high for to be compatible with the effective 
electrical conductivity of the numerical scheme determined 
by the inherent artificial diffusion. Therefore, the ohmic heat flux
might be conceivably two orders of magnitude larger than suggested by
this rough estimate, so that magnetic heating by ohmic dissipation 
must be seriously taken into account. More advanced simulations, taking 
explicit ohmic diffusion into account will clarify this issue. The origin
of current sheets and their microscopic nature is treated in the first
chapter of the second part of these lecture notes.


\section{Future directions}

Continuously increasing power of computational facilities together
with steadily improving computational methods, expand 
the opportunity for numerical simulations. On the one hand, 
more detailed physics can be included in the simulation, on the other 
hand either the computational domain or the
spatial and temporal resolution can be increased. 
In most simulations, especially when the computational domain encompasses
only a small piece of a star, 
boundary conditions play an important role. They convey information on the 
outside world to the physical domain of the simulation. But this 
outside world is often poorly known. In order to acquire experience 
and intuition with respect to the influence of different types of boundary 
condition on the solution, one can implement and run various realizations 
of boundary conditions, which however also requires additional resources 
in computer power and time. Also the initial condition may critically 
determine the solution, for example the net flux and flux density of an 
initial, homogeneous vertical magnetic field. Boundary conditions, therefore, 
remain a hot topic also in future.

Most excitement in carrying out numerical simulations comes 
from the prospect of performing experiments with the object under
investigation: experiments in the numerical laboratory.
Not only that an astrophysical object can be reconstructed and simulated
in the virtual world of the computational astrophysicist. Once in the computer, 
he has the prospect of carrying out experiments as if the celestial body was 
brought to the laboratory.

The following few examples shall illustrate some aspects of this.

\subsection{More detailed physics}

In the solar chromosphere the assumption of LTE (local thermodynamic equilibrium) 
is not valid. Even the assumption of statistical equilibrium in the rate equations 
is not valid because the relaxation time-scale for the ionization of hydrogen 
approaches and surpasses the hydrodynamical time scale in the chromosphere 
\citep{kneer80}. Yet, in order to take time dependent hydrogen 
ionization in a three-dimensional simulation into account, simplifications are 
needed. \citet{leenaarts+wedemeyer06} employ the 
method of fixed radiative rates for a hydrogen model atom with six energy levels 
in the three-dimensional radiation (magneto-)hydrodynamics code \cobold. Thus,
additional to the hydrodynamic equations, they solve the time-dependent rate equations
\begin{equation}
\frac{\partial n_i}{\partial t} + \nabla\cdot(n_i {\bf v}) = 
\sum_{j\ne i}^{n_l} n_j P_{ji} - n_i \sum_{j\ne i}^{n_l} P_{ij}\,,
\end{equation}
with $P_{ij}$ being the sum of collisional and  radiative rate coefficients,
$P_{ij} = C_{ij} + R_{ij}$. The rate coefficients are now local quantities
given a fixed radiation field for each transition, which is obtained from 
one-dimensional test calculations.

Simulations with this approach show that above the height of the classical 
temperature minimum, the non-equilibrium ionization degree is fairly constant 
over time and space at a value set by hot propagating shock 
waves. This is in sharp contrast to results with LTE, where
the ionization degree varies by more than 20 orders of magnitude between 
hot gas immediately behind the shock front and cool regions further away. 
The addition of a hydrogen model atom provides realistic values for hydrogen 
ionization degree and electron density, needed for detailed radiative transfer 
diagnostics. Using this method, \citet{wedemeyer+al07b} have computed 
synthetic maps of the continuum intensity at wavelengths around 1~mm
for a prediction of future high resolution maps that are expected to
be obtained with the Atacama Large Millimeter Array (ALMA).

\subsection{Large box simulations}

\citet{benson+al07} have carried out first simulations with a large
simulation box so as to accommodate a supergranulation cell. They
started a simulation that encompasses a volume of $48 \times 48\times 20$~Mm$^3$ 
using $500^3$ grid cells. With this simulation they hope to find out more
about the large scale motions in the solar convection zone and to carry out 
helioseismic experiments \citep{georgobiani+al07,zhao+al07}. \citet{ustyugov06} carries
out a similar calculation in a box of $(18~\mbox{Mm})^3$ for which he uses
$192\times 192\times 144$ grid points.

\citet{hansteen04} has carried out MHD simulations comprising a vast height 
range from the top layers of the convection zone into the transition region and
the corona. With these simulations they seek to investigate various
chromospheric features such as dynamic fibrils  \citep{hansteen+al06},
mottles, and spicules, which are some of the most important, but also most 
poorly understood, phenomena of the Sun's magnetized outer atmosphere.
But also the transition zone and coronal heating mechanisms are in
the focus of these kinds of ``holistic'' simulations.

\subsection{Improvements in boundary conditions}

Many conventional magnetohydrodynamic simulations of the small-scale solar 
magnetic field assume that the horizontal component of the magnetic field 
vanishes at the top and bottom of the computational domain 
\citep[e.g.][]{weiss+al96,cattaneo+al03,voegler+al05,schaffenberger+al05},
which is a rather stark constraint, especially with respect to 
magnetoacoustic wave propagation and Poynting flux.
Recently, \citet{stein+nordlund06} have introduced an alternative
condition with the possibility of advecting magnetic field across the bottom 
boundary (see also Fig.~\ref{fig_boundary_cond_mag} right). 
Thus, upflows into the computational domain carry horizontal
magnetic field of a prescribed flux density with them, while outflowing
plasma carries whatever magnetic field it instantly has.
With this condition an equilibrium in which equal amounts of
magnetic flux are transported in and out of the computational domain is
approached after some time. It should more faithfully
model the plasma flow across the lower boundary and
it allows for the effect of magnetic pumping 
\citep{tobias+al98,ossendrijver+al02}.

\begin{figure}
\centering
\includegraphics[width=0.48\linewidth]{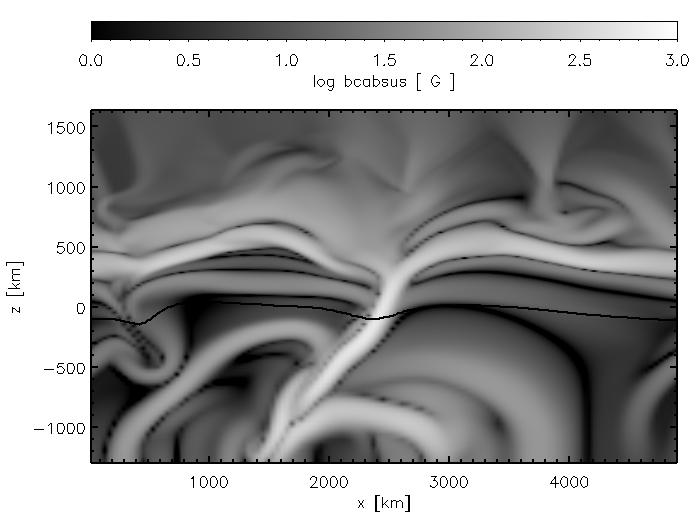}\hfill
\includegraphics[width=0.48\linewidth]{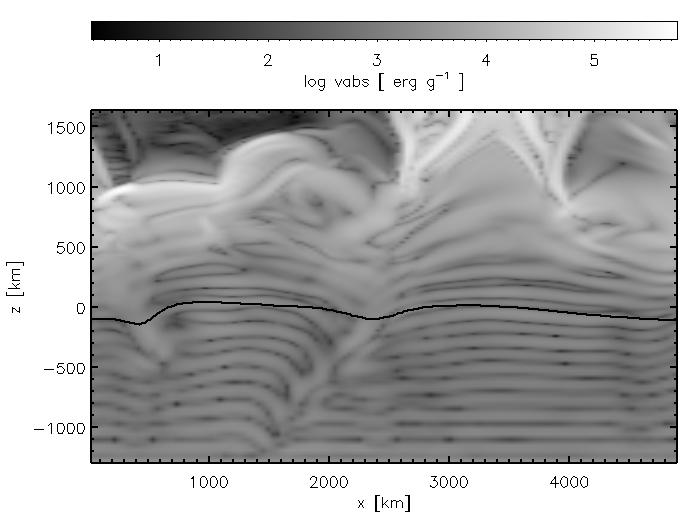}
\caption{
\emph{Left:} Still image of the logarithmic magnetic flux density from a time series for
the instant $t=1368$~s after starting with an initially homogeneous vertical field 
of 1~mT flux density. A strong magnetic flux sheet has formed extending from
$(x,z)\approx (2000,-500)$ to  $(x,z)\approx (2500,0)$.
\emph{Right:} A plane parallel wave with frequency 100 mHz travels through convecting 
plasma into the magnetically structured photosphere and further into the low 
$\beta$ (magnetically dominated) chromosphere. The panel shows the difference 
in absolute velocity between the perturbed and the unperturbed solution 212~s
after the start of the perturbation. The wave becomes strongly refracted
in the low $\beta$ region and at the location of the flux sheet. From \citet{steiner07}.
\label{steiner_fig11}
}
\end{figure}

\subsection{Helioseismic experiment with a magnetically structured atmosphere}

With numerical experiments \citet{steiner+al07} have explored the feasibility of 
using high frequency waves for probing the magnetic fields in the photosphere 
and the chromosphere of the Sun. They track an artificially excited, plane-parallel,
monochromatic wave that propagates through a non-stationary, realistic 
atmosphere, from the convection-zone through the photosphere into the 
magnetically dominated chromosphere, where it gets refracted and reflected.

When comparing the wave travel time between two 
fixed geometrical height levels in the atmosphere (representing the formation
height of two spectral lines) with the topography of the surface of equal magnetic 
and thermal energy density (the magnetic canopy or $\beta=1$ surface) they find good 
correspondence between the two. These numerical experiments support
expectations of \citet{finsterle+al04} that high frequency waves bear 
information on the topography of the `magnetic canopy'.
This simulation exemplifies how a piece of Sun can be made accessible
to virtual experimenting by means of realistic numerical simulation.

\newpage

\section{Part II: Theoretical aspects of magnetic flux tubes}
\label{sect_int_part2}

In this second part we discuss some basic properties of small-scale
magnetic flux tubes. In order to do so, we abstract from the
complexity and peculiarities of observed magnetic flux  
concentration or of simulations as the ones shown in the
first part of these lecture notes. We will consider ideal flux 
tubes with a sharp boundary and often with a circular cross section. 
Although such ideal flux tubes do not exist in the non-ideal 
plasma of the solar photosphere, the approximation is good 
enough for this  idealization to highlight some basic properties
that also pertain to more realistic magnetic flux concentrations.

After definition of an ideal magnetic flux tube we will compute
the current sheet that is associated with the tube surface
and will construct a hydrostatic, vertical flux tube and study
thermodynamic properties of it. We will derive a criterion
for stability with respect to the interchange or fluting.

Supplementary chapters of the lectures that go beyond the scope of 
these notes on flows in a magnetic flux tube and how 
tubes of superequipartition field-strength can be formed and
on wave modes supported by magnetic flux tubes
can be found, e.g., in the book of \citet{stix02}


\section{Concept and properties of magnetic flux tubes}
\label{sect_ft}

A \emph{magnetic flux tube} or \emph{magnetic flux bundle} is 
defined by the surface generated by the set of field lines that 
intersect a simple closed curve.
\begin{figure}[b]
  \includegraphics[width=0.8\textwidth]{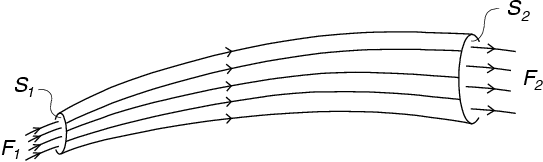}
  \caption{Segment of a magnetic flux tube. After \citet{priest94}.}
\end{figure}
The magnetic flux, crossing a section $S$ of the flux tube is 
given by
\begin{equation}
  \displaystyle F = \int_S {\bf B}\cdot\mbox{d}{\bf S}\,.
\end{equation}
From the divergence free nature of the magnetic field and using 
Gauss's law we obtain
\begin{equation}
  \int\limits_V \nabla\cdot{\bf B}\,\mbox{d}x^3 =
  -\overbrace{\int\limits_{S_1}{\bf B}\,\mbox{d}{\bf s}}^{-F_1}
  +\overbrace{\int\limits_{S_2}{\bf B}\,\mbox{d}{\bf s}}^{F_2}
  +\!\!\!\!\int\limits_{\rm tube\atop surface}\!\!\!\!
   \overbrace{{\bf B}\cdot{\bf\hat{n}}}^0\,\mbox{d}{\sigma} = 0\,,
\end{equation}
where we use that the flux tube surface is by definition parallel
to the magnetic lines of force, hence, ${\bf B}\cdot{\bf\hat{n}}=0$.
As a consequence of $\nabla\cdot{\bf B} = 0$:
\begin{equation}
  F_1 = F_2\,.
  \label{eqn_flux_conservation}
\end{equation}
Hence, like a water hose that keeps mass flux constant, a magnetic 
flux tube conserves the magnetic flux.

\begin{figure}[t]
\centering    
  \includegraphics[width=0.6\textwidth]{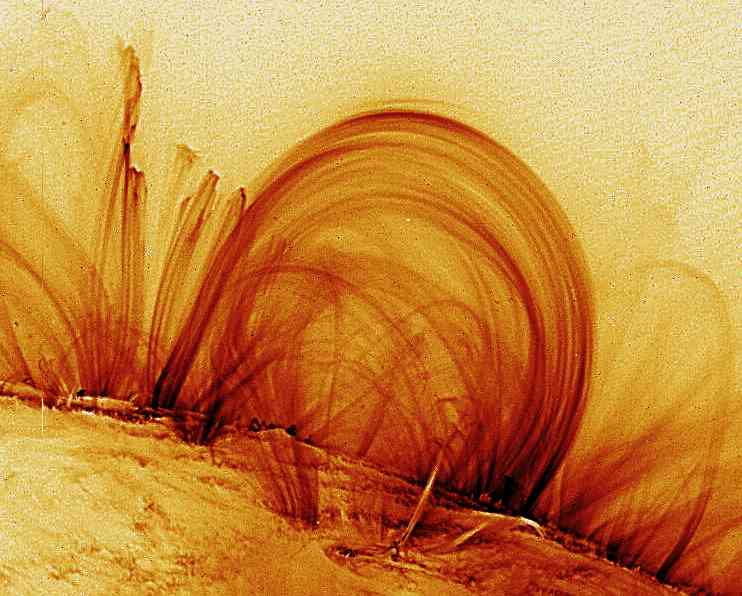}
  \caption{Coronal loops over the solar limb on November 6, 1999 
  taken in the 17.1~nm passband with the Transition Region And Coronal 
  Explorer (TRACE). 
   Cortesy, Stanford-Lokheed Institute for Space Research and NASA.}
   \label{steiner_fig_trace}
\end{figure}

\begin{figure}[b]
\centering
  \includegraphics[width=0.40\textwidth]{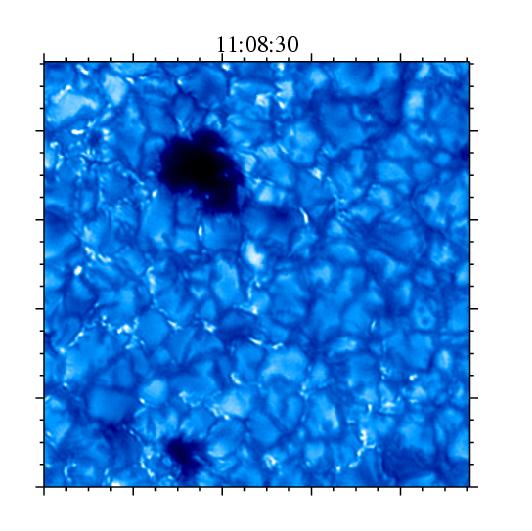}
  \includegraphics[width=0.42\textwidth]{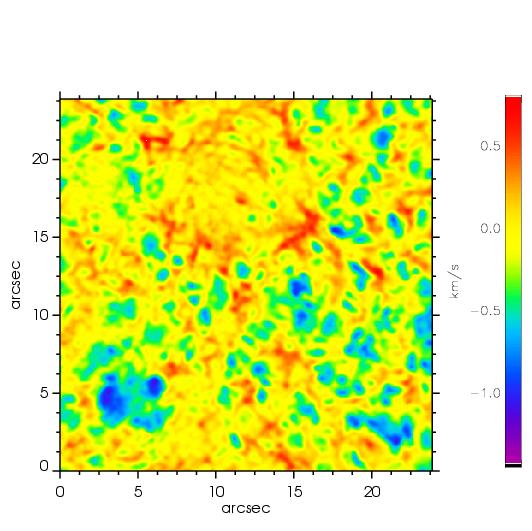}
  \caption{Pore ({left}) and corresponding Dopplergram 
  ({right}). A downdraft (red) exists at the periphery of the pore.
  See internet version for colors.
  Courtesy, Kashia Mikurda, Kiepenheuer-Institut.
  }
  \label{steiner_fig_mikurda1}
\end{figure}

Flux tubes can be considered to be the building blocks of solar
magnetism. However, they must not necessarily be thought of 
as independent isolated structures. Coronal loops for example,
which are close to ideal flux tubes, are embedded in an environment 
that may harbour magnetic fields of similar strength. They just
appear like ``isolated'' flux tubes because the matter within it
happens to be hot enough to emit radiation to become visible in
certain optical passbands. Figure~\ref{steiner_fig_trace} shows
examples of coronal magnetic flux tubes over the solar limb
seen face-on and edge-on.

Pores are another good example of magnetic flux tubes. They
often have a circular like cross section and seem to extend in 
close to radial direction from the convection zone into the
photosphere. Unlike sunspots pores show no penumbra. 
Fig.~\ref{steiner_fig_mikurda1} (left) shows a pore with 
granulation immediately bordering its periphery, giving the 
impression of an ``isolated'' magnetic flux tube embedded in a 
field-free or weak field environment.
Fig.~\ref{steiner_fig_mikurda1} (right) is the corresponding
Dopplergram, showing a downdraft in the periphery of the pore
very much like the downdraft indicated in Fig.~\ref{steiner_fig02}.

Real magnetic flux tubes on the Sun are probably not ideal 
in the sense of Eq.~(\ref{eqn_flux_conservation}) but are more
likely ``leaky'' (\cite{cattaneo+al06}), viz., magnetic lines of force 
do leave and enter the tube keeping Eq.~(\ref{eqn_flux_conservation}) 
only approximately conserved. Also from multidimensional simulations
we know (see chapter on results from multidimensional simulations)
that often field lines of a flux concentrations near the surface disperse again 
in shallow depths so that real flux tubes can be expected to have ``open ends''.
Hence, magnetic flux concentrations at the solar surface may have a 
flux tube character over only a short distance, typically around the level
of optical depth unity.


\subsection{What confines a magnetic flux tube?}

In the following we take a closer look to the flux-tube surface,
where a discontinuity in magnetic field strength and gas pressure
exists. 

\begin{figure}
\centering
  \includegraphics[width=0.30\textwidth]{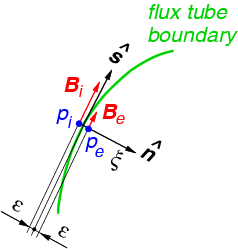}
  \caption{Flux-tube boundary with local coordinates. 
  $\mathbf{\hat{s}}$ is tangential to a field line on the
  tube surface, $\mathbf{\hat{n}}$ perpendicular to it (main normal).
  The field strength at the tube surface immediately inside 
  and outside of it are $\mathbf{B_i}$ and 
  $\mathbf{B_e}$, respectively.
  }
  \label{fig_fs_surface}
\end{figure}

Consider a magnetic flux tube, embedded in a field-free or 
weak-field medium. The discontinuity in magnetic field strength
at the flux-tube boundary can be formally described with the 
step function
$\theta (x)$ with $\theta (x) = 0$ for $x<0$ and
$\theta (x) = 1$ for $x>0$.
In the coordinates of the local frame given by $\bf{\hat{s}}$ and
$\bf{\hat{n}}$, where $\bf{\hat{s}}$ is tangential to the
magnetic field of the flux-tube surface (see Fig.~\ref{fig_fs_surface}),
 $\bf{B}$ is given by
${\bf B} = (0, 0, B_i - [B_i - B_e]\theta(\xi))$.\footnote{A more refined
formulation with an expansion of ${\bf B}_i$ and ${\bf B}_e$ around
the origin of the local frame yields identical results.}
Applying Amp\`{e}re's law: 
\begin{equation}
  \nabla\times{\bf B} = \mu{\bf j}
  \label{eqn_ampere}
\end{equation}
and using $\theta^{\prime}(\xi) = \delta(\xi)$ (Dirac's
$\delta$-distribution) we get:
\begin{equation}
  {\bf j} = \frac{1}{\mu}
  (0, [B]\delta(\xi), 0)\,,
\end{equation}
where $[B] = B_i - B_e$.
Integration over an $\varepsilon$-range and letting
$\varepsilon \rightarrow 0$, leads to the
\emph{current sheet} which flows perpendicular to 
the $\bf{\hat{s}}$-$\bf{\hat{n}}$ plane:
\begin{equation}
  {\bf j^{\ast}} =  \frac{1}{\mu} [B]\,.
  \label{eqn_jstar}
\end{equation}
Notice that this is not a regular current density since
we have carried out a spatial integration in the coordinate $\xi$
in the direction of $\bf{\hat{n}}$ across the flux-tube boundary
so that $\bf j^{\ast}$ has dimension $\mbox{A}\,\mbox{m}^{-1}$.

Thus, we have obtained as a first result that a current sheet
occurs at the flux-tube surface with the current flowing 
perpendicular to the magnetic lines of force. 
Fig.~\ref{steiner_fig10} nicely demonstrates how current sheets may
occur in the actual solar atmosphere. The characteristic
double layers in the panel corresponding to 90~km above 
the mean surface of optical depth unity originate from the
current sheet that flows on either side (surface with a sharp
gradient in $B$) of the magnetic flux sheets that are located 
in the intergranular lanes.

Next, we consider
the force balance starting from the momentum equation
\begin{equation}
  \rho\displaystyle\frac{\mbox{D}{\bf v}}{\mbox{D}t} = -\nabla p
  +({\bf j}\times{\bf B}) + \rho {\bf g},
\end{equation}
where the middle term on the right hand side is the Lorentz force
and ${\bf g}$ the gravitational acceleration at the
solar surface (optical depth $\tau_{500\,\mathrm{nm}} = 1$). Again using
Amp\`{e}re's law, Eq.~ (\ref{eqn_ampere}), we obtain for the 
\emph{static case}
\begin{equation}
  \nabla p = \frac{1}{\mu}(({\nabla}\times{\bf B})\times{\bf B})
  + \rho {\bf g}
\end{equation}
and further using the vector identity
$(\nabla\times{\bf B})\times{\bf B} = ({\bf B}\cdot\nabla){\bf B}
-\displaystyle\frac{1}{2}\nabla({\bf B}\cdot{\bf B}),
$
\begin{equation}
\nabla p = -\nabla(\displaystyle\frac{B^2}{2\mu}) + \rho {\bf g}
+\displaystyle\frac{1}{\mu}({\bf B}\cdot\nabla){\bf B}\,.
  \label{eqn_mag_tensions}
\end{equation}
Now we decompose the last term into a component
parallel (${\bf\hat{s}}$) and perpendicular (main normal ${\bf\hat{n}}$)
to a surface field line.
From ${\bf B} = B\,{\bf\hat{s}}$ follows:
${\bf B}\cdot\nabla = B\cdot{\bf\hat{s}}\cdot\nabla = 
B(\partial/\partial s)$
and further
\begin{equation}
({\bf B}\cdot\nabla){\bf B} = 
(B\displaystyle\frac{\partial}{\partial s}){\bf B} =
 B\displaystyle\frac{\partial}{\partial s}(B{\bf\hat{s}}) =
 B\displaystyle\frac{\partial B}{\partial s}{\bf\hat{s}} + 
 B^2\displaystyle\frac{\partial{\bf\hat{s}}}{\partial s} = 
 \displaystyle\frac{\partial}{\partial s}
(\displaystyle\frac{B^2}{2}){\bf\hat{s}} + 
 B^2\displaystyle\frac{\bf\hat{n}}{R_c}\;,
 \label{eqn_curvature}
\end{equation}
where $R_c$ is the curvature radius of the field line.
Thus, we obtain:
\begin{equation}
\nabla p = -\nabla(\displaystyle\frac{B^2}{2\mu}) + \rho {\bf g}
+\displaystyle\frac{\partial}{\partial s}(\frac{B^2}{2\mu})\;{\bf\hat{s}}
+\displaystyle\frac{B^2}{\mu}\frac{\bf\hat{n}}{R_c}\;,
\end{equation}
where we note that the term ${B^2}/{2\mu}$ acts like a gas pressure,
called the \emph{magnetic pressure} \citep[see][chap.~2.7]{priest82}.
Multiplication of this equation with ${\bf\hat{n}}$ yields the force
balance perpendicular to the tube surface:
\begin{equation}
\displaystyle\frac{\partial p}{\partial n} = 
-\displaystyle\frac{\partial}{\partial n}
(\displaystyle\frac{B^2}{2\mu}) + 
\rho {\bf g}\cdot{\bf\hat{n}} +
\displaystyle\frac{B^2}{\mu}\displaystyle\frac{1}{R_c}\,
\end{equation}
and integration over a small interval $[-\epsilon, \epsilon]$
across the tube surface gives:
\begin{displaymath}
  \underbrace{\int\limits_{p_i}^{p_e}\mbox{d}p}_{p_e - p_i} = 
 \, \underbrace{-\!\!\!\!\int\limits_{B_i^2/2\mu}^{B_e^2/2\mu}
     \mbox{d}(\frac{B^2}{2\mu})}_{-\frac{1}{2\mu}(B_e^2 - B_i^2)} +
  \rho {\bf g}\cdot{\bf\hat{n}} 2\epsilon +
  \frac{{\bar{B}}^2}{\mu R_c} 2\epsilon\;.
\end{displaymath}
Letting $\varepsilon \rightarrow 0$, we finally obtain:
\begin{equation}
  p_e + \frac{B_e^2}{2\mu} = p_i + \frac{B_i^2}{2\mu}\;.
  \label{eqn_pbalance}
\end{equation}
Thus, we have obtain as a second result that a discontinuity
in pressure occurs at the flux-tube surface so that the Lorentz
force is balanced by a corresponding pressure force. It is
important to realize that Eq.~(\ref{eqn_pbalance}) was obtained
from basic principles without using the 
``thin flux-tube approximation''. Eq.~(\ref{eqn_pbalance})
expresses that the total pressure consisting of gas pressure
plus magnetic pressure is constant across the flux tube surface.
This results holds generally for tangential magnetic discontinuities in
a plasma. Note however, that indices $i$ and $e$ refer to
locations in the immediate vicinity of the discontinuity.


\subsection{A microscopic picture of the current sheet}

\begin{figure}
\begin{minipage}{0.47\textwidth}
  \includegraphics[width=\textwidth]{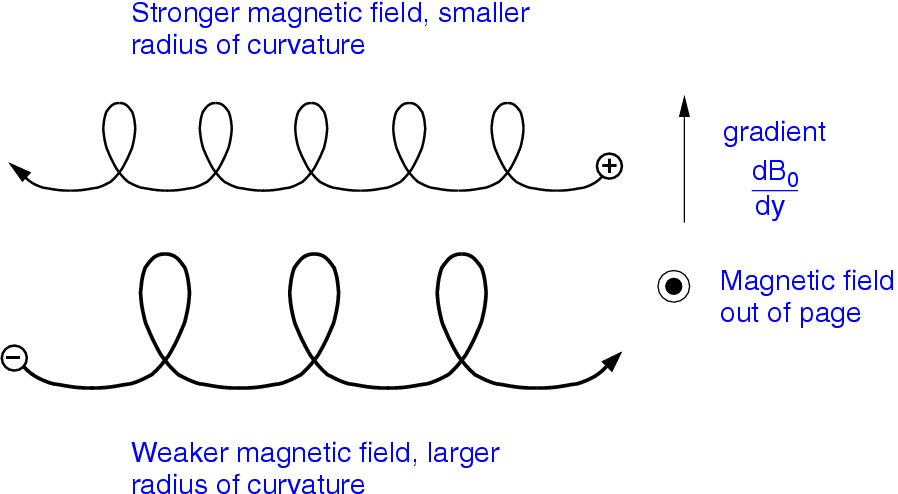}
\end{minipage}\hspace*{0.05\textwidth}
\begin{minipage}{0.47\textwidth}
  \includegraphics[width=\textwidth]{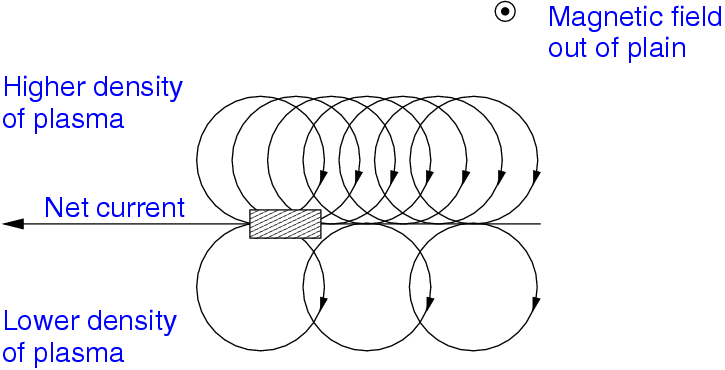}
\end{minipage}
\caption{\emph{Left:} Particle motion in a magnetic field with a transverse gradient showing
grad $B$ drift motion (adapted from \citet{krall+trivelpiece73}). \emph{Right:} Particle 
orbits in a plasma with a sharp density gradient resulting in a net drift current (right). 
}
\label{fig_drift01}
\end{figure}

It may seem surprising that an electrical current can be confined
to a thin sheet in a bulk of plasma with high conductivity everywhere in 
space. This subsection is intended to sketch the microscopic processes
that lead to this current sheet. Consider first the situation in which
we have a uniformly directed magnetic field with a spatial gradient in 
field strength as plotted in Fig.~\ref{fig_drift01} (left). The field
is directed perpendicular to the drawing plane pointing out of it and
increasing in strength from bottom to top. A negatively charged particle
that gyrates in this field will unabatedly drift to the right because 
the Lorentz force, ${\bf F}_L = q ({\bf v}\times {\bf B})$, is
not constant along the path of the particle but increases as it moves 
into higher field strength leading to a reduction in curvature radius.
Likewise, a positively charged particle will drift to the left.
Hence, electrons and ions in a magnetic field with a transverse
gradient in flux density show a \emph{gradient B drift motion},
leading to a net electrical current. 

Next we consider in Fig.~\ref{fig_drift01} (right) the situation
of a uniform magnetic field pointing out of the 
drawing plane. Instead of the gradient in field strength we
have a contact discontinuity in density with a lower density
in the lower half of the figure and a higher density in the upper 
half. In each half, positively charged particle will gyrate in a 
circle in the sense like indicated in Fig.~\ref{fig_drift01} (right),
however, there are much more gyrating particles in the upper half
than in the lower half because of the difference in particle density.
Considering now the hatched control volume, which straddles the contact
discontinuity and which is much smaller that the gyration radius, we
can readily see that much more charged particles run to the left
than to the right, leading to a net positively charged current to 
the left and, correspondingly, a net negatively charged current to 
the right. Due to the imbalance of gyrating particles, an electric
current results without a net transport of charges. No net current
results for a control volume in either the upper or the lower
half if it does not include the discontinuity.

\begin{figure}[b]
  \includegraphics[width=0.6\textwidth]{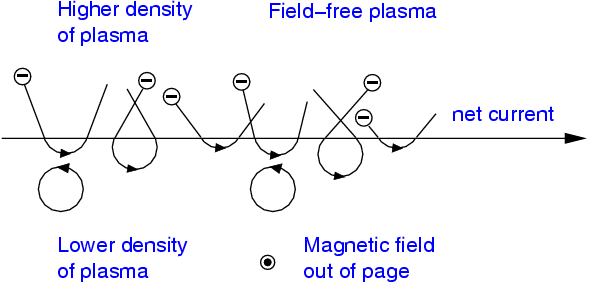}
\caption{Particle orbits in a plasma with a sharp gradient in density and
magnetic field strength as it occurs on the surface of a magnetic flux tube.
It results a net drift current confined to the flux-tube surface.
}
\label{fig_drift03}
\end{figure}
Finally, Fig.~\ref{fig_drift03} sketches the situation at the
boundary of a flux tube that is thought to be embedded in a
field free environment. Charged particles
in the field-free plasma move on straight free paths but start
to gyrate as soon as they enter the magnetic field of the flux
tube, where negatively charged particles will drift to the right
for the situation depicted in Fig.~\ref{fig_drift03}. However,
this drift current will not be cancelled by gyrating particles
within the flux tube because of the reduced particle number-density 
there. Hence, we obtain a net drift current with a net transport of
charges.


\section{The equations for a hydrostatic flux tube}

In this section we derive the equations that govern the magnetic
structure of a vertical, axisymmetric, untwisted magnetic flux tube embedded
in a gravitationally stratified, field-free atmosphere. Although
such a perfectly symmetric object is an idealization that does
probably not occur in reality, the basic properties of the solutions 
to these equations can be recovered in the more complex situations
that we know from numerical simulations in a three-dimensional
environment. The system of the magnetohydrostatic equations to
be solved is
\begin{eqnarray}
  0 &=& -\nabla p + \rho{\bf g} + {\bf j}\times{\bf B}\;,\label{eqn_hsmoment}   \\
  \nabla\times {\bf B} &=& \mu {\bf j}\;,\label{eqn_hsampere} \\
  \nabla\cdot{\bf B} &=& 0 \label{eqn_hsmaxwell}\;.
\end{eqnarray}
We decompose Eq.~(\ref{eqn_hsmoment}) in components parallel and 
perpendicular to the magnetic field. Thus, multiplying  
Eq.~(\ref{eqn_hsmoment}) with ${\bf B}$ gives
\begin{equation}
  {\bf B}\cdot(\nabla p - \rho {\bf g}) = 0 
  \label{eqn_parallel}
\end{equation}
and taking the cross product of  Eq.~(\ref{eqn_hsmoment}) with 
{\bf B} gives
\begin{equation}
  {\bf j} = \frac{1}{B^2}{\bf B}\times (\nabla p - \rho {\bf g})\;,
  \label{eqn_perpendicular}
\end{equation}
using that $\; {\bf B}\times({\bf j}\times{\bf B}) = 
B^2{\bf j}-({\bf B}\cdot{\bf j})\cdot{\bf B} = B^2{\bf j}\;$
for an untwisted axisymmetric field.

Next we reform Amp\`{e}re's law, Eq.~(\ref{eqn_hsampere}).
Using cylindrical coordinates and Eq.~(\ref{eqn_hsmaxwell})
we first have
\begin{equation}
{\bf B}\! =\! \nabla\times{\bf A}\! =\! \left(
\displaystyle\frac{1}{r}\frac{\partial A_z}{\partial\phi}-
\displaystyle\frac{\partial A_{\phi}}{\partial z}\;,\;
\displaystyle\frac{\partial A_r}{\partial z}-
\displaystyle\frac{\partial A_z}{\partial r}\;,\;
\displaystyle\frac{1}{r}(\frac{\partial}{\partial r}(r A_{\phi})-
\displaystyle\frac{\partial A_r}{\partial\phi})\right)\;,
\end{equation}
which reduces to
\begin{equation}
{\bf B} = \left(
-\displaystyle\frac{\partial A_{\phi}}{\partial z}\;,\;
0\;,\;
\displaystyle\frac{1}{r}
\displaystyle\frac{\partial}{\partial r}(r A_{\phi})\right)
\end{equation}
for an axisymmetric flux tube without twist. With the
stream function $\Psi:=rA_{\phi}$ we can write
\begin{equation}
{\bf B} =\left(
-\displaystyle\frac{1}{r}\displaystyle\frac{\partial\Psi}{\partial z}\;,\;
0\;,\;
\displaystyle\frac{1}{r}\displaystyle\frac{\partial\Psi}{\partial r}\right)
\end{equation}
and further, taking the curl of ${\bf B}$
\begin{equation}
\nabla\times{\bf B} = \left(0\;,\;
-\displaystyle\frac{1}{r}\displaystyle\frac{\partial^2 \Psi}{\partial z^2}+
\displaystyle\frac{1}{r^2}\displaystyle\frac{\partial\Psi}{\partial r}-
\displaystyle\frac{1}{r}\displaystyle\frac{\partial^2 \Psi}{\partial r^2}\;,0\right)\;,
\end{equation}
so that Amp\`{e}res law for the $\phi$-component (the $r$- and
$z$-components being zero), becomes 
(Grad-Shafranov):
\begin{equation}
  \displaystyle
   \frac{\partial^2 \Psi}{\partial r^2}
  -\frac{1}{r}\displaystyle\frac{\partial\Psi}{\partial r}
  +\frac{\partial^2 \Psi}{\partial z^2}
  = -\mu rj_{\phi}\;.
  \label{eqn_grad-shafranov}
\end{equation}
This is a second order elliptic, inhomogeneous, partial differential
equation. It is non-linear because $j_{\phi}$ depends on $\Psi$ again
as we will see further down in the text.
The magnetic field components can be recovered from the stream function
$\Psi$:
\begin{equation}
   B_r =-\frac{1}{r}\frac{\partial\Psi}{\partial z}\;,\qquad
   B_z = \frac{1}{r}\frac{\partial\Psi}{\partial r}\;.
   \label{eqn_bofpsi}
\end{equation}
Note that contours of $\Psi$ are field lines of the 
system. 

\begin{figure}
  \includegraphics[width=0.2\textwidth]{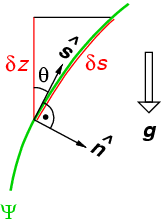}
  \caption{Magnetic field line $\Psi$ with local coordinate frame.
  $\mathbf{\hat{s}}$ is tangential to the field line, $\mathbf{\hat{n}}$ 
  is the main normal.
  }
  \label{fig_decompose}
\end{figure}

Now we consider Eq.~(\ref{eqn_parallel}) and Fig.~\ref{fig_decompose}.
If gravity acts parallel to the $z$-axis in the negative direction 
and $s$ measures the distance along a magnetic field line inclined 
at an angle $\theta$ to the vertical direction, we get from
Eq.~(\ref{eqn_parallel})
\begin{equation}
  \frac{\mbox{d}p}{\mbox{d}s} + \rho g \cos\theta = 0\,.
\end{equation}
Substituting the ideal gas equation for the density $\rho$
we obtain
\begin{equation}
  \displaystyle\frac{\mbox{d}p}{p} = 
 -\displaystyle\frac{\bar{m}g}{k_B T}\mbox{d}z\,.
\end{equation}
$H={k_BT}/(\bar{m}g)$ is a length scale, called the pressure scale 
height, because the gas pressure drops by a factor $\mbox{e}$ over this
distance in the outward direction. Note that $H$ is proportional
to the temperature. $\bar{m}$ is the mean molecular weight. Integration
from a reference height $z=0$ to an arbitrary height $z$ along
a magnetic line of force labeled with $\Psi$ gives
\begin{equation}
  p = p_0(\Psi) \exp\!\!\left[-\int\limits_{\Psi,0}^{\Psi,\,z}
  \frac{\mbox{d}z^{\,\prime}}{H(T(\Psi,z^{\,\prime}))}\right]\;.
  \label{eqn_hydrostatics}
\end{equation}
This equation expresses that
hydrostatic equilibrium holds along magnetic field lines.
Picking a field line, we obtain with this equation the 
gas pressure, $p(z=0)$, exerted by the mass column along this 
field line.

Having made use of the longitudinal component of Eq.~(\ref{eqn_hsmoment})
we now turn to its perpendicular component, Eq.~(\ref{eqn_perpendicular}). Using
cylindrical coordinates
${\bf B} = (B_r\;,\;0\;,\;B_z)$ and 
$\nabla p = (\partial p/\partial r\;,\;0\;,\;\partial p/\partial z)$,
Eq.~(\ref{eqn_perpendicular}) becomes
\begin{equation}
  \displaystyle
  j_{\phi} = \frac{1}{B^2}\left[B_z\frac{\partial p}{\partial r} -
  B_r(\frac{\partial p}{\partial z} + \frac{p}{H})\right]\;.
\end{equation}
From Eq.~(\ref{eqn_hydrostatics}) we obtain
\begin{equation}
   \displaystyle\frac{\partial p}{\partial z} = 
  -\displaystyle\frac{p}{H} +
   \left. \frac{\partial p}{\partial \Psi}\right|_z 
   \frac{\partial \Psi}{\partial z}
\qquad \mbox{and}\qquad
   \displaystyle\frac{\partial p}{\partial r} = 
   \left. \frac{\partial p}{\partial \Psi}\right|_z 
   \frac{\partial \Psi}{\partial r}
\end{equation}
so that the above equation for $j_{\phi}$ reduces to
\begin{equation}
  \displaystyle
  j_{\phi} = r \left. \frac{\partial p}{\partial \Psi}\right|_z \,.
  \label{eqn_volume_current}
\end{equation}
At the surface of the tube the magnetic field will,
in general, be discontinuous, resulting in a sheet
current there. Using 
\begin{equation}
  p_e - p_i = \frac{1}{2\mu}(B_i^2  - B_e^2) = 
  \frac{1}{2\mu}(B_i - B_e)(B_i + B_e)
\end{equation}
and Eq.~(\ref{eqn_jstar}) we finally get
\begin{equation}
  \displaystyle
  j^{\ast}_{\phi} = \frac{2(p_e - p_i)}{B_i + B_e}\,.
  \label{eqn_sheet_current}
\end{equation}

The procedure for computing the magnetic structure of a 
vertically oriented, axisymmetric flux tube without twist
is the following:

\begin{list}{-}{\setlength{\parsep}{0.0mm}}
\item[(i)]   Specify an initial magnetic configuration, e.g.,
             with the help of the thin flux-tube approximation.
             This initial flux tube shall be embedded in a given
             plane-parallel atmosphere that is thought to remain 
             unchanged.
\item[(ii)]  Calculate the gas pressure everywhere in space using
             hydrostatics along magnetic lines of force, 
             Eq.~(\ref{eqn_hydrostatics}). Since we do not solve
             an energy equation here, the pressure scale height as
             a function of field line and height, $H(\Psi,z)$, 
             must also be specified. An acceptable guess would be
             that $H$ is independent of $\Psi$ and identical to
             the scale height in the surrounding field-free
             atmosphere, which is approximately fulfilled if
             the temperature is constant in planes parallel to
             the solar surface. The pressure and pressure scale height 
             in the surrounding field-free atmosphere is known from
             the initial, embedding atmosphere.
\item[(iii)] Evaluate volume and current sheet using
             Eqs.~(\ref{eqn_volume_current}) and 
             (\ref{eqn_sheet_current}). Eq.~(\ref{eqn_volume_current})
             applies to regions of continuous gas pressure,
             Eq.~(\ref{eqn_sheet_current}) to locations of 
             discontinuous gas pressure and field strength like
             the surface of the magnetic flux tube.
\item[(iv)]  Solve the Grad-Shafranov equation, Eq.~(\ref{eqn_grad-shafranov}),
             with appropriate boundary conditions. This leads to
             a new field configuration, $\Psi(r,z)$, that must not 
             necessarily be identical to the initial magnetic configuration.
             Hence, one must go back to (ii) and iterate (ii) to (iv)
             until convergence has been achieved.
\end{list}

\begin{figure}
  \includegraphics[width=0.4\textwidth]{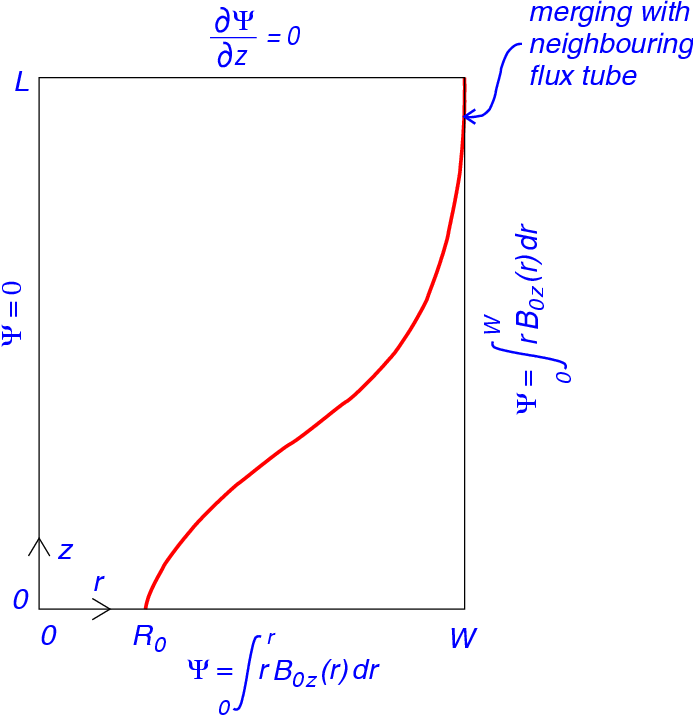}
  \caption{Boundary conditions for solving the elliptic partial differential
  equation, Eq.~(\ref{eqn_grad-shafranov}).
  }
  \label{fig_bc_grad-shafranov}
\end{figure}

Boundary conditions need to be specified for solving 
Eq.~(\ref{eqn_grad-shafranov}). A plausible choice is shown in
Fig.~\ref{fig_bc_grad-shafranov}.
The field line with $\Psi = 0$ that marks the left boundary
of the computational domain coincides with the symmetry axis
of the flux tube. At the bottom boundary the vertical component
of the magnetic field as a function of radial distance is prescribed,
which fixes $\Psi(r,z=0)$.  The boundary on the right hand side 
is thought to separate the flux tube from fields of neighbouring
flux tubes. We can think of it as a separating field line of
constant $\Psi$. At the top we assume that the magnetic field
merges with the magnetic field of similar neighbouring flux tubes
leading into a homogeneous, vertical magnetic field with vanishing
horizontal component. Formally, this condition is obtained by
setting $\partial\Psi/\partial z = -r B_r = 0$.
The width $W$ of the computational domain 
determines the filling factor $f$, viz., the fraction of the
area at $z=0$ that is occupied with magnetic field:
\begin{equation}
  f = W^2/R_0^2\;.
\end{equation}
Eq.~(\ref{eqn_grad-shafranov}) together with these boundary 
conditions constitutes a \emph{free boundary problem} because the
boundary of the magnetic flux tube is not known from the beginning
but is part of the solution.

Instead of using a rectangular computational domain, \citet{pizzo90}
uses a ``body-fitted'' nonorthogonal coordinate system to map the 
physical domain of the flux tube into a unit square computational 
domain. A multigrid elliptic solver is used at each iteration stage 
for solving the Grad-Shafranov equation. 
\citet{fiedler+cally90} use a similar mesh in which contours
of constant $\Psi$ (field lines) constitute one coordinate,
the normalized arc length along field lines the second one.
\citet{jahn89} and \citet{jahn+schmidt94} use a similar
method for the construction of sunspot models.

In case of horizontal temperature
equilibrium, $T_i(z) = T_e(z) = T(z)$, we have (neglecting any 
horizontal variation in ionization degree) $H_i(z) = H_e(z) = H(z)$. 
Then, 
\begin{equation}
\displaystyle p_i(z) = p_{0\,i} \exp\!\!\left[-\int\limits_{0}^{z}
  \frac{\mbox{d}z^{\,\prime}}{H(T(z^{\,\prime}))}\right]\quad 
 \displaystyle p_e(z) = p_{0\,e} \exp\!\!\left[-\int\limits_{0}^{z}
  \frac{\mbox{d}z^{\,\prime}}{H(T(z^{\,\prime}))}\right]\,,
\end{equation}
from which follows that $p_i(z) < p_e(z)\, \forall z $ assuming
that $p_i$ does not depend on radius (thin tube approximation)
because $p_{0e} - p_{0i} = B_0^2/2\mu > 0$. Since $T_i(z) = T_e(z)$
it follows for the densities that $\rho_i(z) < \rho_e(z)\, \forall z $.
Since above conditions can be expected to hold very well in the
photosphere we can say that photospheric flux tubes are rarefied,
one also says ``partially evacuated''.

As a most simple case we explore in the following a flux tube
with a constant axial field strength and gas pressure at the base 
level, $z_0$, where the radius is $R_0$. Then

\begin{equation}
  \Psi(r,z_0) = \left\{ \begin{array}{l@{\quad\mbox{for}\quad}l}
  B_{z\,0}r^2/2   & r\le R_0\\
  B_{z\,0}R_0^2/2 & r\ge R_0
  \end{array}\right.
\end{equation}
from Eq.~(\ref{eqn_bofpsi}) and 
\begin{equation}
  p_{i\,0} = p_{e\,0} - \frac{B_{z\,0}^2 + B_{r\,0}^2(R_0)}{2\mu}
\end{equation}
from the requirement of pressure balance, Eq.~(\ref{eqn_pbalance}).
We note, that the final $B_{r\,0}(R_0)$ is not known from the very
beginning so that the boundary condition for the pressure needs
to be adjusted in the course of the iteration.

If we further assume that the temperature
at a given hight level is constant, then the gas pressure is 
constant too, so that the volume current is
\begin{equation}
  j_{\phi} = r \left. \frac{\partial p}{\partial \Psi}\right|_z = 0
\end{equation}
and we have a potential field \emph{inside} the flux tube. 
However, there remains a current sheet at the tube surface:
\begin{equation}
  j^{\ast}_{\phi} = \frac{2(p_e - p_i)}{B_i + B_e}\;.
\end{equation}

\begin{figure}
  \includegraphics[width=0.40\textwidth]{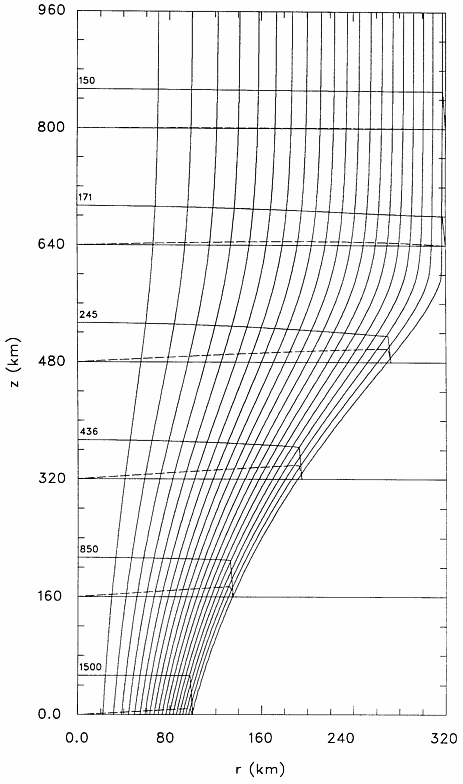}
  \caption{Field lines and cross sections for a cylindrically symmetric, vertical
  flux tube with uniform pressure and axial field strength at the base, carrying
  a current sheet at the surface. Superimposed on the figure are plots of
  the radial variation of $B_r$ and $B_z$ at different heights, normalized to the 
  value of $B_z$ at the axis, each indicated in gauss ($10^4$~T). Note that the
  vertical scale is compressed relative to the horizontal scale by a factor 1.6.
  From \citet{steiner+al86}.
  }
  \label{ft_1500}
\end{figure}

Fig.~\ref{ft_1500} shows the corresponding solution
for a flux tube with a field strength of 0.15~T
and a radius of 100~km at the base of the photosphere 
($\tau_{500\, \mathrm{nm}} = 1$). The figure shows representative
lines of force of the magnetic field. We recall that
the left hand boundary coincides with the symmetry axis of
the cylindrical tube. The ring areas between two field
lines contain equal amounts of magnetic flux.

Superimposed on the field-line plot are plots of the radial 
variation of $B_r$ (dashed curve) and of $B_z$ (solid curve)
at different heights, both normalized to the value of $B_z$ at the
axis, indicated in gauss ($10^4$ T). The flux tube merges with 
the field of neighbouring flux tubes at a height of $\approx 500$~km. 
The filling factor is $f = (R_0/W)^2 = 0.1$.

Due to the exponentially decreasing gas pressure in the gravitationally
stratified atmosphere, the magnetic flux tube must expand with height
because of pressure balance, Eq.~(\ref{eqn_pbalance}), and
flux conservation resulting from $\nabla\, {\bf B} = 0$.
Even though, the axial field strength is constant with radius at
the reference height $z=0$, this is no more the case higher
up in the photosphere, where $B_z(r)$ decreases towards the
flux-tube boundary. This variation of the axial field strength
with flux-tube radius may become much stronger than in the
case of Fig.~\ref{ft_1500} depending on the expansion rate of
the flux tube, which in turn depends on the combination of
internal to external atmosphere. If $H_i = H_e$ as was assumed
for the flux tube of Fig.~\ref{ft_1500} it can be shown with
the help of the thin flux-tube approximation that
it takes about four pressure scale heights for the flux-tube
radius to expand by a factor of $\mbox{e}$ and about two 
pressure scale heights for the axial field strength to decrease 
by a factor of $\mbox{e}$.

Introducing the scalar function $G=rB_{\phi}$ one can treat
the case of a twisted axisymmetric flux tube. The Grad-Shafranov
equation then becomes
\begin{equation}
   \left(
   \frac{\partial G}{\partial z}\;,\;
   \frac{\partial^2 \Psi}{\partial r^2}
  -\frac{1}{r}\displaystyle\frac{\partial\Psi}{\partial r}
  +\frac{\partial^2 \Psi}{\partial z^2}\;,\;
  -\frac{\partial G}{\partial r}
   \right)
  = -\mu r\,{\bf j}\;.
\end{equation}
If $G = G(\Psi)$ (torque-free condition) we need only solve
the $\phi$-component, for which:
\begin{equation}
    j_{\phi} = r \left. \frac{\partial p}{\partial \Psi}\right|_z 
    + \frac{1}{\mu r}G\frac{\partial G}{\partial \Psi}\;.
\end{equation}
In the absence of an external magnetic field, the magnitude
of the current sheet remains the same as before:
$ |{\bf j^{\ast}}| = {2(p_e - p_i)}/{B_i}$, directed
perpendicular to the field lines at the surface and, hence,
no longer purely azimuthal. The $\phi$-component of the
current sheet is:
\begin{equation}
  j_{\phi}^{\ast} = \frac{2(p_e - p_i)}{B_i}
  \sqrt{1-(\frac{B_{\phi}}{B_i})^2}\;.
\end{equation}
The solution procedure remains the same as before. Without even
carrying out a stability analysis it can be shown that there exists an 
upper limit for twisting the flux tube beyond which, there is
no magnetohydrostatics possible because the magnetic tension
at the tube surface can no longer be counterbalanced by gas
and magnetic pressure in the tube interior.


\subsection{The magnetic canopy}

The expansion rate of hydrostatic flux tubes critically depends on
the combination of external to internal atmosphere. This is demonstrated
by the following ideal model in which we assume constant but different
pressure scale heights, $H_e$ and $H_i$  in the atmospheres external 
and internal to the magnetic flux tube, respectively. If 
$H_i > H_e$, which is equivalent to a hot flux-tube atmosphere,
$T_i > T_e$, a critical hight, $z_{\mathrm{crit}}$, ensues,
where the flux tube must strongly expand in the horizontal direction. 

\begin{figure}[h]
  \includegraphics[width=0.35\textwidth]{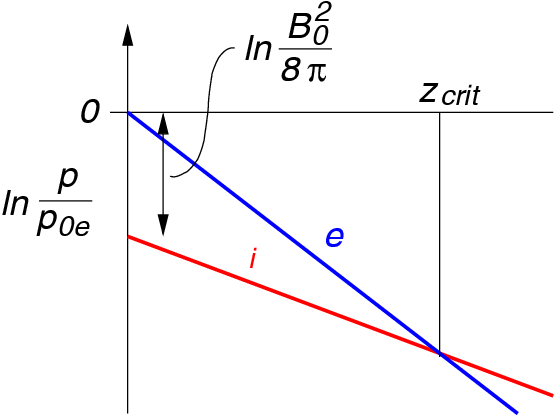}
  \caption{Gas pressure as a function of height within (i) and external (e)
  to the flux tube. At the critical height, $z_{\mathrm{crit}}$, the magnetic
  field is forced to expand in the horizontal direction.
  }
  \label{fig_zcrit}
\end{figure}

This situation is illustrated in Fig.~\ref{fig_zcrit}, which shows the
logarithmic gas pressure normalized to the external gas pressure
at the reference height, $z_0$, as a function of height in the
atmosphere. Initially, the internal gas pressure, $p_i$, is smaller
than the gas pressure in the external atmosphere, $p_e$,
for to accommodate the magnetic flux tube with magnetic
pressure $p_{\mathrm{mag}} = p_e - p_i$. But at the critical height,
$z_{\mathrm{crit}}$, the internal gas pressure approaches the external
gas pressure so that the magnetic field (magnetic pressure) at the
tube surface must vanish, which causes the sudden horizontal 
expansion.

Generally, the pressure stratification in the internal and external
atmosphere are given by
\begin{equation}
   p_{i,e}(z) = p_{0\,i,e} \exp\!\!\displaystyle
   \left(-\displaystyle\int\limits_{0}^{z}
   \displaystyle\frac{\mathrm{d}z^{\,\prime}}{H_{i,e}(z^{\,\prime})}\right).
\end{equation}
With $p_e(z_{\mathrm{crit}}) = p_i(z_{\mathrm{crit}})$ and assuming
constant pressure scale heights, we obtain
\begin{equation}
   \int\limits_0^{z_{\mathrm{crit}}}(\frac{1}{H_e} - \frac{1}{H_i})
   \mathrm{d}z = \ln \frac{p_{0\,e}}{p_{0\,i}}
   \quad\stackrel{H = \mathrm{const.\rule[-5pt]{0.0pt}{10pt}}}{\Rightarrow}\quad
   z_{\mathrm{crit}} = \frac{H_i\, H_e}{H_i-H_e}\ln
   \frac{p_{0\,e}}{p_{0\,i}} =
   \frac{H_i\, H_e}{H_i-H_e}\ln
   \frac{\beta_{0} + 1}{\beta_{0}}\;,
\end{equation}
where we use the plasma beta, $\beta_0 = p_{0\,i}/p_{\mathrm{mag}}$,
the ration of gas to magnetic pressure at the reference height $z_0$.

\begin{figure}
  \includegraphics[width=0.5\textwidth]{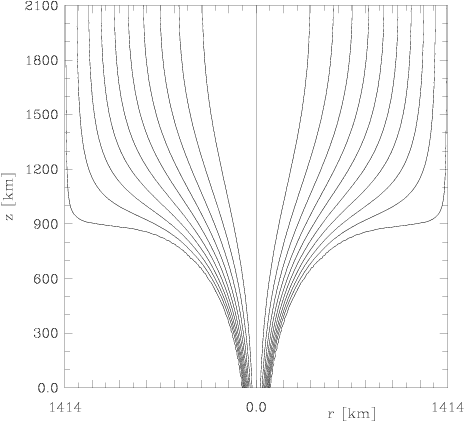}
  \caption{Flux tube with a standard quiet sun atmosphere embedded in a
  radiative equilibrium atmosphere without a chromospheric temperature rise. 
  The field-strength at the base ($z=0$) is 0.15~T. The field lines 
  spread into a horizontally extending canopy field at a height of 900\,km.
  }
  \label{fig_ft_canopy}
\end{figure}

As an example, Fig.~\ref{fig_ft_canopy} shows magnetic lines of force
of a rotationally symmetric, vertical flux tube that harbours a
standard quiet sun atmosphere (C$^{\prime}$ of \citet{maltby+al86}, Table 11). 
The external atmosphere is an atmosphere in radiative equilibrium 
showing no chromospheric temperature rise \cite{anderson89}.
Therefore, above the temperature minimum of the internal atmosphere, at a 
height of approximately $z=500$\,km, it becomes distinctly  cooler than the
flux-tube atmosphere with the consequence that $p_i$ rapidly
approaches $p_e$ so that the field lines spread into the horizontal
direction at a height of about 900\,km forming a so called \emph{canopy field}.
The strength of the canopy field is weak -- in fact, the magnetic pressure
must vanish at the boundary to the canopy as we have seen. Therefore,
the plasma $\beta$ is very inhomogeneous within a tube like that of 
Fig.~\ref{fig_ft_canopy} and the surface of $\beta=1$ spreads much less
dramatic with height than the peripheral field lines might suggest.

\begin{figure}
  \includegraphics[width=0.6\textwidth]{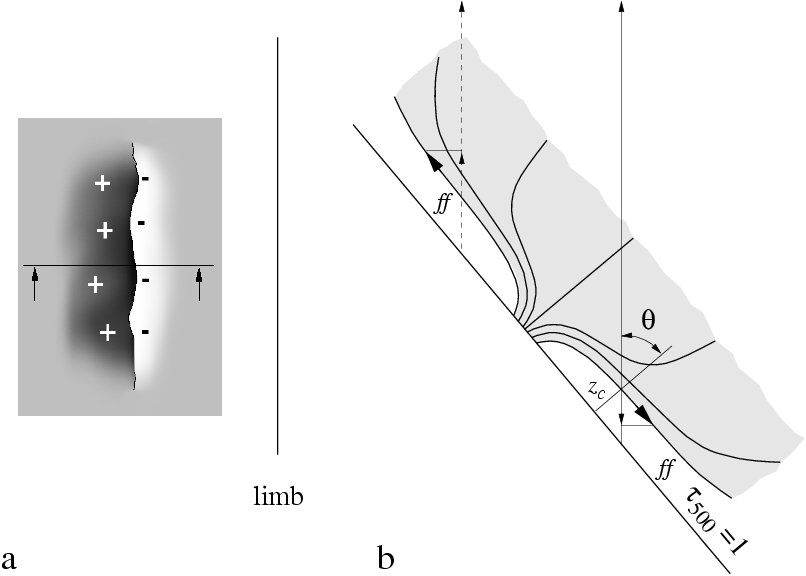}
  \caption{Sketch of a chromospheric magnetogram near the solar limb (a) 
  and the corresponding magnetic configuration (b). The base of the
  magnetic canopy is located at a height $z_{\mathrm{c}}$ above the solar
  surface. From \citet{steiner2000}.
  }
  \label{fig_mg}
\end{figure}

Canopy fields also occur in the superpenumbra of sunspots or in the
surroundings of network magnetic fields. The latter can be directly
observed in limb magnetograms of chromospheric spectral lines. This is
illustrated in the sketch of Fig.~\ref{fig_mg}, which shows an expanding
magnetic flux tube that should be thought to represent a network
magnetic flux concentration. The solar surface ($\tau_{500} = 1$) is
inclined with respect to the horizontal to indicate observation near the
solar limb. The hatched region is filled with magnetic field of the flux 
concentration, while {\it ff} indicates the field-free region. 

Lines of sight passing through the central network field concentration 
receive a magnetic field component toward the observer and so do lines 
of sight passing through the canopy field to the left (disk center) side of 
the network (dashed line of sight). The canopy field to the right 
(limbward) side of the network, however, gives rise to a line of 
sight component of opposite direction (solid line of sight).
Thus, the observed magnetic polarity inverses across a line that 
coincides with the limbward edge of a unipolar magnetic network 
field as is sketched in the corresponding magnetogram to the left.

Indeed it was observed that chromospheric magnetograms of a network 
region that is unipolar in magnetograms of photospheric
lines (that form below the canopy) show near the limb a 
\emph{fringe pattern} in polarity \citep{giovanelli80}. 
This fringe pattern can be easily 
understood when we imagine the sketch of  Fig.~\ref{fig_mg} (right) to be
continued to a sequence of neighbouring unipolar flux concentrations,
each giving rise to a chromospheric magnetogram like the one
of Fig.~\ref{fig_mg} (left). From magnetograms taken in the line 
of Mg I b2 (517.3~nm) \citet{giovanelli+jones82} derived canopy 
heights of 600-1000~km in quiet-Sun regions and as low as 200~km 
in active regions.


\section{Magnetic flux tube in radiative equilibrium}

There are two basic modes of energy transport in the solar photosphere 
and convection zone: radiative and convective. When in a stationary state 
all of the energy is transported by radiation, we have 
\emph{radiative equilibrium}, conversely, in the case of pure convective transport
we have \emph{convective equilibrium}. In a stationary transport process, the 
frequency distribution of the radiation, or the portioning of energy between 
the radiative and the convective mode of transfer, may be altered; but
the energy flux as a whole is rigorously conserved. Formally, this is 
expressed by
\begin{equation}
  \nabla\cdot{\bf F}_{\mathrm{tot}} = 0\,,\qquad \mbox{where}\qquad
  {\bf F}_{\mathrm{tot}} = {\bf F}_{\mathrm{rad}} + {\bf F}_{\mathrm{conv}}\;,
\end{equation}
assuming that no energy transport other than radiative and convective
be relevant, as is usually assumed in mixing length theory.

For the two limiting cases mentioned above we have
\begin{eqnarray}
\mbox{radiative  equilibrium:}   & \nabla\cdot{\bf F}_{\mathrm{rad}} = 0 
                                                 & {\bf F}_{\mathrm{con}} = 0\;,\\
\mbox{convective equilibrium:} & \nabla\cdot{\bf F}_{\mathrm{con}} = 0 
                                                 & {\bf F}_{\mathrm{rad}} = 0\;.
\end{eqnarray} 
In the solar photosphere radiative energy transfer by large
prevails so that $\nabla\cdot{\bf F}_{\mathrm{rad}}  = 0$ is
a good approximation. With $I({\bf r}, {\bf\hat{n}}, \nu)$
being the radiative intensity with frequency $\nu$ propagating
in direction ${\bf\hat{n}}$ at location ${\bf r}$ in 
three-dimensional  space, the total radiative flux is given by
\begin{equation}
  {\bf F}_{\mathrm{rad}} =
  \int\limits_{4\pi}\!\int\limits_0^{\infty}
  I({\bf r}, {\bf\hat{n}}, \nu) {\bf\hat{n}}
  \,\mathrm{d}\nu\,\mathrm{d}\omega\;.
\end{equation}
$I({\bf r}, {\bf\hat{n}}, \nu)$ has dimension 
W\,m$^{-2}$\,hz$^{-1}$\,sr$^{-1}$ and correspondingly
${\bf F}_{\mathrm{rad}}$ has dimension W\,m$^{-2}$.
The radiation field follows as a solution of the radiative
transfer equation in three-dimensional space \citep{mihalas_stat}
\begin{equation}
  ({\bf\hat{n}}\cdot\nabla)I({\bf r}, {\bf\hat{n}}, \nu) =
  \eta({\bf r}, \nu) - \kappa({\bf r}, \nu)
  I({\bf r}, {\bf\hat{n}}, \nu)\;.
  \label{eqn_radtra}
\end{equation}
The emissivity $\eta({\bf r}, \nu)$ is given by 
$\eta({\bf r}, \nu) = \kappa({\bf r}, \nu) S({\bf r}, \nu)$
with $S$ being the source function. If $S({\bf r})$
and $ \kappa({\bf r})$ are known, Eq.~(\ref{eqn_radtra})
is an ordinary first order differential equation whose
solution is called the \emph{formal solution} of the radiative
transfer equation.

From these equations we obtain:
\begin{eqnarray}
  \nabla\cdot{\bf F}_{\mathrm{rad}} &=& 
  \int\limits_{4\pi}\!\int\limits_0^{\infty}
  ({\bf\hat{n}}\cdot\nabla) I({\bf r}, {\bf\hat{n}}, \nu) 
  \,\mathrm{d}\nu\,\mathrm{d}\omega\\ 
  &= &
  \int\limits_{4\pi}\!\int\limits_0^{\infty}
  (\kappa({\bf r}, \nu) S({\bf r}, \nu) - \kappa({\bf r}, \nu)
  I({\bf r}, {\bf\hat{n}}, \nu) \stackrel{!}{=} 0\;.
\end{eqnarray}
With
\begin{equation}
  J({\bf r}, \nu) = \displaystyle\frac{1}{4\pi}\int\limits_{4\pi}
  I({\bf r}, {\bf\hat{n}}, \nu) \,\mathrm{d}\omega
  \label{eqn_mean_intensity}
\end{equation}
being the mean intensity, we finally obtain the 
constraint equation for radiative equilibrium:
\begin{equation}
  \int\limits_0^{\infty}
  \kappa({\bf r}, \nu) S({\bf r}, \nu)\,\mathrm{d}\nu = 
  \int\limits_0^{\infty}
  \kappa({\bf r}, \nu) J({\bf r}, \nu)\,\mathrm{d}\nu\;. 
  \label{eqn_constraint}
\end{equation}
In general we do not know the temperature distribution
$T({\bf r})$ that satisfies radiative equilibrium. If
we start with a guess $T^{(0)}({\bf r})$ for which we
have calculated the correct source function
$S({\bf r}, \nu)$, we will find that the
constraint equation for radiative equilibrium is not
satisfied. It is therefore necessary to iteratively adjust
$T({\bf r})$ until the requirement of radiative balance is
satisfied.

Assume a given temperature distribution $T({\bf r})$ of
the magnetohydrostatic configuration for which we also know
the pressure $p({\bf r})$ and density $\rho({\bf r})$. This
allows us to derive LTE-values for the opacities $\kappa({\bf r}, \nu)$.
We then are able to compute the radiation field everywhere in space
${\bf r}$ and for all frequencies $\nu$ by evaluation of the 
formal solution of the radiative transfer equation
\begin{equation}
  J_{\nu}({\bf r}) = {\Lambda}_{\nu}
  ({\bf r}, {\bf r^{\prime}}) B_{\nu}({\bf r^{\prime}})
  + G_{\nu}({\bf r})\;,
  \label{eqn_formal}
\end{equation}
where ${\Lambda}_{\nu}$ is the integral operator
which adds the intensities at ${\bf r}$ caused by emission
at all the points ${\bf r^{\prime}}$ in the considered
computational domain, and where $G_{\nu}$ is the transmitted
mean intensity due to the incident radiation field into
this domain. We also use strict local thermodynamic equilibrium 
(LTE) ($S_{\nu} = B_{\nu}$), although a scattering component could 
be included in the solution method that follows. Note
that ${\Lambda}_{\nu}$ together with $ G_{\nu}$ is just a 
short-hand for the integration, Eq.~\ref{eqn_mean_intensity}, 
and the formal solution of the radiative transfer equation,
Eq.~(\ref{eqn_radtra}). In practice we can think of 
${\Lambda}_{\nu}$ as a matrix operator acting on a 
vector of source values given at discrete spatial locations.

\begin{figure}
  \includegraphics[width=0.42\textwidth]{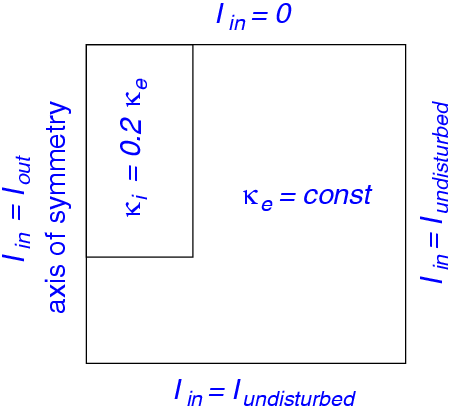}
  \hspace*{0.05\textwidth}
  \includegraphics[width=0.42\textwidth]{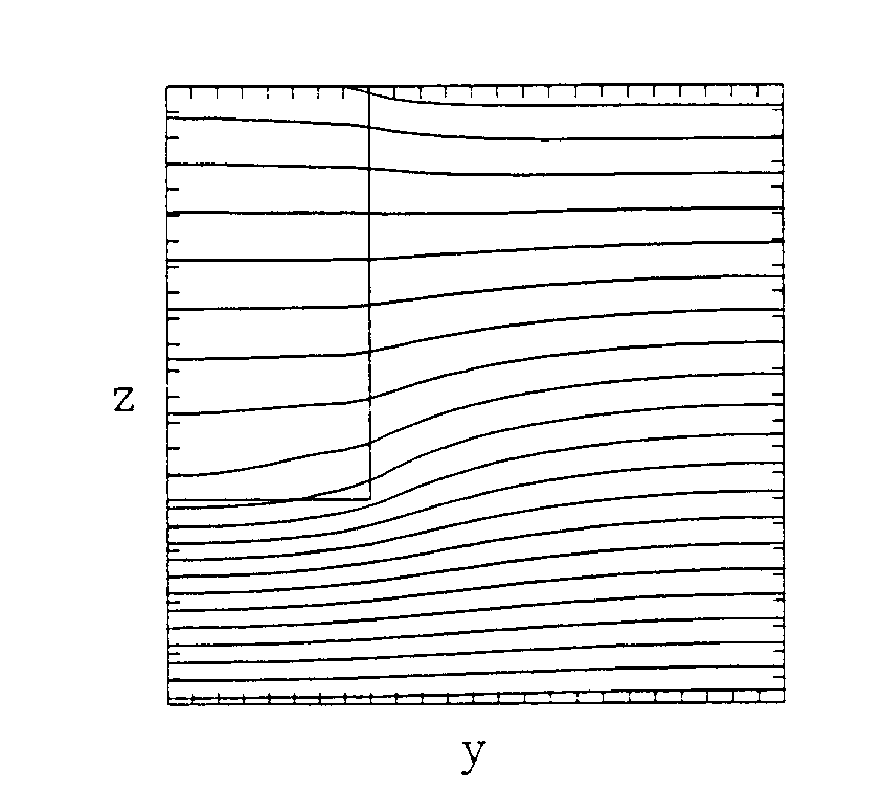}
\caption{Two-dimensionally structured atmosphere in radiative equilibrium. 
\emph{Left:} Radiative boundary conditions (for determining $G({\bf r})$ of 
Eq.~\ref{eqn_formal}) and distribution of opacity. \emph{Right:} Isotherms
of the radiative equilibrium solution. From \citet{steiner90}.
}
  \label{fig_grey_2d}
\end{figure}

Defining the integral operator ${\cal{K}}$ so that
$
  {\cal{K}}\phi:= 
  \int_0^{\infty}\kappa_{\nu}\phi\mathrm{d}\nu\;,
$
where $\phi$ is a scalar function, we can write the constraint 
equation for radiative equilibrium, Eq.~(\ref{eqn_constraint}), as:
\begin{equation}
  {\cal{K}} B_{\nu} = 
  {\cal{K}} {\Lambda}_{\nu} B_{\nu}
  + {\cal{K}} G_{\nu}\,.
\end{equation}
Given an initial temperature distribution,
$T^{(0)}({\bf r})$, this equation is in general not 
satisfied. But we can compute a correction 
$\Delta T({\bf r})$ such that
\begin{equation}
  {\cal{K}} B_{\nu}(T^{(0)}+\Delta T) = 
  {\cal{K}} {\Lambda}_{\nu} B_{\nu}(T^{(0)})
  + {\cal{K}} G_{\nu}
\end{equation}
is satisfied. Expanding $B_{\nu}(T+\Delta T)\approx B_{\nu}(T) + 
(\partial B_{\nu}/\partial T)|_T\Delta T$, we can compute the 
temperature correction:
\begin{equation}
  \Delta T = \frac{{\cal{K}}
  ({\Lambda}_{\nu} - {\bf I}) B_{\nu}(T)
  + {\cal{K}} G_{\nu}}
  {{\cal{K}}(\partial B_{\nu}/\partial T)|_T}\,.
\end{equation}

\begin{figure}
  \includegraphics[width=0.6\textwidth]{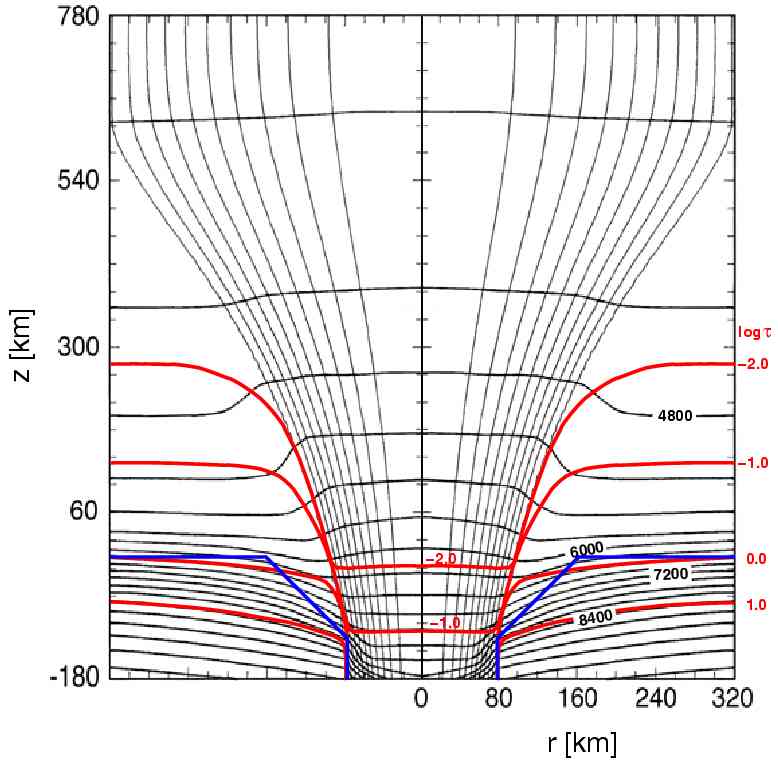}
\caption{Radiative equilibrium solution for a cylindrically symmetric,
vertical magnetic flux tube with a field strength of $\approx 0.15$~T
at $z = 0$. The horizontally running contours are isotherms labeled
with the temperature in K. The curves labeled in the right margin of 
the figure are contours of constant optical depth for vertical lines of 
sight. The temperature in the convectively unstable region of the
indicated pentagonal torus was prescribed. Adapted from
\citet{steiner+stenflo90}.
}
  \label{fig_ftradeq}
\end{figure}

For most practical purposes this iterative scheme is
very slow. The next better scheme to use is a Jacobi-like
iteration, in radiative transfer known as 
\emph{accelerated ${\Lambda}$-iteration}\,:
\begin{equation}
  \Delta T = \frac{{\cal{K}}
  ({\Lambda}_{\nu} - {\bf I}) B_{\nu}(T)
  + {\cal{K}} G_{\nu}}
  {{\cal{K}}(1 - \lambda^{\ast}_{\nu})
  (\partial B_{\nu}/\partial T)|_T}\;,
\end{equation}
where $\lambda^{\ast}_{\nu}$ is the diagonal element of
the matrix representing ${\Lambda}_{\nu}({\bf r})$.
Note that $\Delta T = \Delta T({\bf r})$ and that the above
equation must be evaluated for each location ${\bf r}$.
Since the formal solution of the transfer equation must
be computed in each iteration at each location ${\bf r}$, 
an efficient solver is needed, which is given by the 
so called method of short characteristics 
\citep{kunasz+auer88,kunasz+olson88}
and by exploiting the symmetry properties in the case
of a rotationally symmetric flux tube \citep{steiner93}.

Fig.~\ref{fig_grey_2d} is a striking example of a two-dimensionally 
structured atmosphere in radiative equilibrium. The panel to the left 
shows the boundary condition (for determining $G({\bf r})$ of 
Eq.~\ref{eqn_formal}) and the distribution of opacity. The rectangle 
in the upper left corner may represent a magnetic flux sheet
with a rarified atmosphere and correspondingly reduced opacity. 
The incident radiation on the bottom
and on the right side of the computational domain is computed
from the undisturbed, plane-parallel atmosphere with 
$\kappa = \kappa_e = \mbox{const}$. The left side is an axis of symmetry of the
configuration. The top boundary has no incident radiation.
The panel to the right shows the isotherms of the radiative
equilibrium solution obtained with the above given solution
procedure. While in deep layers the rarified region 
is cooler than it is in the unperturbed surrounding atmosphere,
the top layers are hotter.

\begin{figure}
  \includegraphics[width=0.45\textwidth]{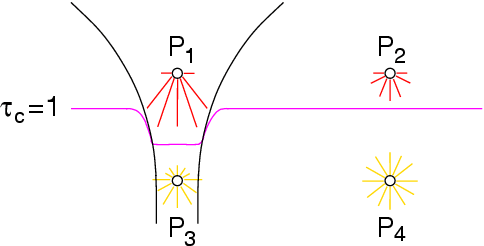}
\caption{Irradiation on point $P_1$ located on the flux-tube axis
above the level of $\tau_{\mathrm{c}} = 1$ and on point $P_2$ located
at the same geometrical height but laterally separated from the
flux tube in the undisturbed atmosphere. Adapted from \citet{steiner_phd}.
}
  \label{fig_irrad}
\end{figure}

This bahaviour can be recovered when computing the radiative
equilibrium solution of a magnetic flux tube, like the one shown
in Fig.~\ref{fig_ftradeq}. It shows isotherms  together with contours
of constant optical depth for vertical lines of sight, labeled in
the right margin of the figure. The vertically running curves 
represent lines of force of the magnetic field. The frequency
integration in the constraint equation for radiative equilibrium,
Eq.~(\ref{eqn_constraint}) was done using opacity distribution
functions of \citet{kurucz79}, thus, taking line blanketing effects
into account \citep{steiner+stenflo90}. The
height scale on the left hand side originates at the optical depth
$\tau_{500\,\mathrm{nm}}=1$ in the surrounding, unperturbed atmosphere,
from which follows that the computational domain extends into the
top surface layers of the convection zone, where the assumption
of radiative equilibrium breaks down. Therefore, the temperature
structure within the indicated pentagonal torus below $z=0$ and outside
the magnetic field was prescribed and kept constant when
computing the radiative equilibrium solution. Resort to this 
stratagem obviously reveals the limits of the radiative equilibrium 
approach. Yet, this solution highlights a typical property of
flux-tube atmospheres that can be recovered in fully self-consistent
simulations: the temperature along the axis of symmetry increases 
in the deep photospheric layers less strong with depth than in the
surrounding, unperturbed atmosphere, hence, the flux-tube atmosphere
has a ``flat temperature gradient''. This is a consequence of the
``radiative channeling effect'' by which the photospheric layers
of the flux tube get heated through influx of radiation from the
optically denser surroundings. This change in temperature gradient
has far reaching consequences for the formation of spectral lines
within the magnetic flux tube. Generally, spectral lines tend
to become weaker (line weakening) with the consequence that 
filtergrams of spectral regions with a high density of spectral 
lines show magnetic elements of particularly high contrast 
(e.g., G-band filtergrams).

Why is the temperature in the photospheric layers of the magnetic 
flux tube by about 200~K higher than in the surrounding atmosphere?
Consider two points $P_1$ and $P_2$ located at equal geometrical
height, $P_1$ on the flux tube axis, $P_2$ far away from the flux tube
in the unperturbed external atmosphere, as is illustrated in
Fig.~\ref{fig_irrad}. $P_1$ receives a higher intensity
of radiation from the flux tube's hot walls (radiative channeling effect)
as compared to $P_2$. Both points receive equal amounts of low
radiative intensity from the top. Therefore, the mean intensity is higher 
in $P_1$ than in  $P_2$, $J_1 > J_2$. Since in radiative equilibrium
$J = S = \int B_{\nu}\mathrm{d}\nu \propto T^4$, $T_1 > T_2$.
The opposite behaviour occurs for points $P_3$ and $P_4$ as
long as the assumption of radiative equilibrium holds. However, the
true reason for the reduced temperature in the optically thick layers of the
flux tube is the inhibition of convective energy transport by the
magnetic field.


\section{Interchange instability of magnetic flux tubes}

\begin{figure}[h]
  \includegraphics[width=0.65\textwidth]{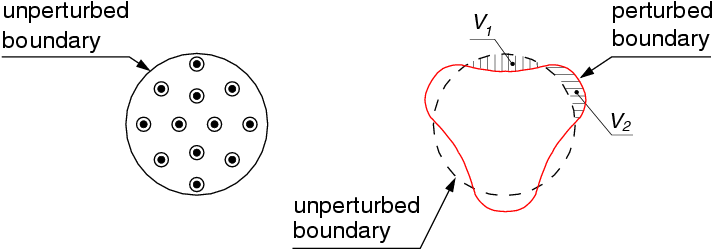}
\caption{\emph{Left:} Cross section of an unperturbed cylindrical magnetic 
  flux tube. \emph{Right:} Same view after perturbation showing the initial
  perturbation of the flutes. Adapted from \citet{krall+trivelpiece73}.
}
  \label{fig_flute}
\end{figure}

Consider a section perpendicular to the axis of a straight flux-tube
as is sketched in Fig.~\ref{fig_flute}. A perturbation of the flux-tube 
boundary in such a way that the volumes $V_1$ and $V_2$ are equal, 
does not change the total energy of the configuration as equal amounts
of magnetic and thermal energy are exchanged.
To zeroth order, the perturbation also does not change the energy 
state of a flux tube embedded in a stratified medium.

The instability that may evolve from this perturbation
is called \emph{interchange instability}, as the
magnetic field and gas of volume 1 is interchanged with the
magnetic field and gas of volume 2. It is also called 
\emph{flute instability} because of the shape
of the perturbed surface (like the vertical parallel grooves on a
classical architectural column, called flute).

\begin{figure}
  \includegraphics[width=0.38\textwidth]{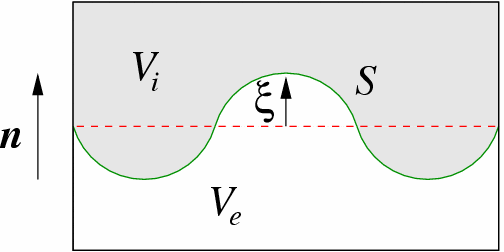}
\caption{Distorted surface $S$ displaced from the equilibrium
position (dashed line). The unit vector ${\bf n}$ is normal to
the unperturbed surface and the section is normal to the 
magnetic field lines close to the tube surface. 
Adapted from \citet{meyer+al77}.
}
  \label{fig_flute2}
\end{figure}

Consider a small flux-tube section with a small perturbation ${\xi}$ (Fig.~\ref{fig_flute2}),
where the grey shaded area, $V_i$, be the magnetic flux tube while $V_e$ is field-free. 
${\bf n}$ is the surface normal pointing out of the field-free plasma.
In order that there is a net restoring force on the displaced
surface we must, for the indicated displacement ${\xi}$, have 
\begin{equation}
  p_{0\,i_{\,\mathrm{tot}}} 
  + \delta p_{i_{\,\mathrm{tot}}} >  
  p_{0\,e} + \delta p_e\quad \Rightarrow\quad
  \delta p_{i_{\,\mathrm{tot}}} > \delta p_e
\end{equation}
since for the equilibrium configuration 
$ p_{0\,i_{\,\mathrm{tot}}} = p_{0\,e}$.
From this follows the condition for stability:
\begin{equation}
  |{\xi}\cdot {\bf n}|{\bf n}\cdot\nabla p_e <
  |{\xi}\cdot {\bf n}|{\bf n}\cdot\nabla
  (p_i + \frac{B^2}{2\mu})\;.
  \label{eqn_bernstein}
\end{equation}
This criterion also follows from a more general energy principle
due to \citet{bernstein+al58}. It is both, necessary and sufficient
for stability.

Within the flux tube we have 
\begin{equation}
  \nabla p_i + \nabla(\displaystyle\frac{B^2}{2\mu})= \rho_i\, {\bf g}
  +\displaystyle\frac{1}{\mu}({\bf B}\cdot\nabla){\bf B}
  \label{eqn_flute0}
\end{equation}
as was derived in connection with Eq.~(\ref{eqn_mag_tensions}).
In the external atmosphere we have hydrostatic equilibrium
\begin{equation}
  \nabla p_e = \rho_e {\bf g}\;.
  \label{eqn_flute1}
\end{equation}
Using these equations in the previously derived stability criterion, 
Eq.~(\ref{eqn_bernstein}), we obtain:
\begin{equation}
  {\bf n}\cdot\left[\frac{1}{\mu}({\bf B}\cdot\nabla){\bf B} - 
  (\rho_e - \rho_i){\bf g} \right] > 0\;.
  \label{eqn_flute2}
\end{equation}
From Eqs.~(\ref{eqn_flute0}) and (\ref{eqn_flute1}) we have
\begin{equation}
  \frac{1}{\mu}({\bf B}\cdot\nabla){\bf B} - 
  (\rho_e - \rho_i){\bf g} = \nabla(
  \underbrace{p_i + \frac{B^2}{2\mu} - 
  p_e}_{=0 \mbox{~on~} S})\;,
  \label{eqn_flute3}
\end{equation}
which means that the bracketed vector in Eq.~(\ref{eqn_flute2}) is 
parallel to ${\bf n}$ on the surface $S$ of the flux tube.

Therefore, gravity can be eliminated taking the horizontal component
of this vector. If ${\bf h}$ is a horizontal vector pointing out
of the flux tube into the field-free plasma we get
\begin{equation}
  {\bf h}\cdot[({\bf B}\cdot\nabla){\bf B}] < 0\:.
  \label{eqn_flute4}
\end{equation}
This means that along any field line in $S$ the magnitude of
the component of ${\bf B}$ in any fixed outward horizontal direction 
must decrease as a function of height in the atmosphere for to
have stability. For an untwisted axisymmetric flux tube this criterion 
can be expressed in cylinder coordinates as:
\begin{equation}
  \left.\frac{\mbox{d}B_r}{\mbox{d}z}\right|_S < 0\;.
  \label{eqn_flute_stability1}
\end{equation}

From 
${\bf n}\cdot ({\bf B}\cdot\nabla){\bf B} = 
-\displaystyle\frac{B^2}{R_c}$ (see Eq.~\ref{eqn_curvature}),
where $R_c$ is the curvature radius 
of the surface in axial direction, results another useful form 
of the stability criterion for an axisymmetric untwisted flux tube:
\begin{equation}
  -\frac{B^2}{\mu R_c} + 
  (\rho_e - \rho_i)g\sin\chi > 0\;.
  \label{eqn_flute_stability2}
\end{equation}
$\chi$ is the inclination of $S$ with respect to the vertical direction.

Using a realistic model atmosphere, \citet{meyer+al77} came to the 
conclusion that only flux tubes with magnetic flux $\Phi > 10^{11}$~Wb
are stable against the flute instability. Sunspots and pores
have flux in excess of $10^{11}$~Wb. With typical values of
$R = 100$~km and $B = 0.1$~T resulting in 
$\Phi \approx 3\cdot 10^{9}$~Wb, small-scale magnetic flux
concentrations in plage and network regions are liable to 
the flute instability!

Indeed, from time sequences of G-band-bright-point images
(e.g., \cite{berger+al04,vandervoort05})
one gets the impression that magnetic elements are subject to
fluting (see also the discussion in the first chapter of part I). 
On the other hand their typical life time of 6--8
minutes is still in excess of the crossing time for
Alfv\'en or sound waves of about 100~s. Also there are bright
points that undergo continual fragmentation and merging in a 
relatively stable location,  persisting over several hours.


\begin{theacknowledgments}
I would like to thank the director H.~M.~Antia and the convener
K.~E.~Rangarajan for having invited me to, and for their excellent 
organization of the 2006 Solar Physics Winter School at Kodaikanal 
Solar Observatory. I also thank the director of the Indian Institute
of Astrophysics (IAA) S.~S.~Hasan for his kind hospitality during my stay
at the IAA in Bangalore. This work was supported by the German Academic
Exchange Service (DAAD), grant D/05/57687, and the Indian Department of
Science \& Technology (DST), grant DST/INT/DAAD/P146/2006. 
\end{theacknowledgments}


\def\aj{AJ}%
\def\Astronomische Nachrichten Supplement{Astron. Nachr. Suppl.}%
\def\araa{ARA\&A}%
\def\apj{ApJ}%
\def\apjl{ApJ}%
\def\apjs{ApJS}%
\def\ao{Appl.~Opt.}%
\def\apss{Ap\&SS}%
\def\aap{A\&A}%
\def\aapr{A\&A~Rev.}%
\def\aaps{A\&AS}%
\def\azh{AZh}%
\def\baas{BAAS}%
\def\jrasc{JRASC}%
\def\memras{MmRAS}%
\def\mnras{MNRAS}%
\def\pra{Phys.~Rev.~A}%
\def\prb{Phys.~Rev.~B}%
\def\prc{Phys.~Rev.~C}%
\def\prd{Phys.~Rev.~D}%
\def\pre{Phys.~Rev.~E}%
\def\prl{Phys.~Rev.~Lett.}%
\def\pasp{PASP}%
\def\pasj{PASJ}%
\def\qjras{QJRAS}%
\def\skytel{S\&T}%
\def\solphys{Solar~Phys.}%
\def\sovast{Soviet~Ast.}%
\def\ssr{Space~Sci.~Rev.}%
\def\zap{ZAp}%
\def\nat{Nature}%
\def\iaucirc{IAU~Circ.}%
\def\aplett{Astrophys.~Lett.}%
\def\apspr{Astrophys.~Space~Phys.~Res.}%
\def\bain{Bull.~Astron.~Inst.~Netherlands}%
\def\fcp{Fund.~Cosmic~Phys.}%
\def\gca{Geochim.~Cosmochim.~Acta}%
\def\grl{Geophys.~Res.~Lett.}%
\def\jcp{J.~Chem.~Phys.}%
\def\jgr{J.~Geophys.~Res.}%
\def\jqsrt{J.~Quant.~Spec.~Radiat.~Transf.}%
\def\memsai{Mem.~Soc.~Astron.~Italiana}%
\def\nphysa{Nucl.~Phys.~A}%
\def\physrep{Phys.~Rep.}%
\def\physscr{Phys.~Scr}%
\def\planss{Planet.~Space~Sci.}%
\def\procspie{Proc.~SPIE}%
\let\astap=\aap
\let\apjlett=\apjl
\let\apjsupp=\apjs
\let\applopt=\ao

\bibliographystyle{aipproc}   

\bibliography{steiner}

\IfFileExists{\jobname.bbl}{}
 {\typeout{}
  \typeout{******************************************}
  \typeout{** Please run "bibtex \jobname" to optain}
  \typeout{** the bibliography and then re-run LaTeX}
  \typeout{** twice to fix the references!}
  \typeout{******************************************}
  \typeout{}
 }

\end{document}